\documentclass[aps,pre,twocolumn,superscriptaddress,showpacs,showkeys]{revtex4-1}
\usepackage{graphicx} 
\usepackage{hyperref}                   
\usepackage{amsmath,amssymb}            
\usepackage{epstopdf}
\usepackage{subcaption}					
\usepackage{float}
\usepackage{color}
\definecolor{orange}{rgb}{1,0.5,0}
\definecolor{brown}{rgb}{0.65, 0.16, 0.16}
\definecolor{phlox}{rgb}{0.87, 0.0, 1.0}

\begin{document}
\title{Self-organized criticality in cumulus clouds}

\author{M. N. Najafi}
\affiliation{Department of Physics, University of Mohaghegh Ardabili, P.O. Box 179, Ardabil, Iran}
\email{morteza.nattagh@gmail.com}
\affiliation{Computational Physics, IfB, ETH Zurich, Stefano-Franscini-Platz 3, CH-8093 Zurich, Switzerland}

\author{J. Cheraghalizadeh*}
\affiliation{Department of Physics, University of Mohaghegh Ardabili, P.O. Box 179, Ardabil, Iran}
\email{jafarcheraghalizadeh@gmail.com}

\author{H. J. Herrmann}
\affiliation{ESPCI, CNRS UMR 7636 - Laboratoire PMMH, 75005 Paris (France)}
\email{hans@ifb.baug.ethz.ch}
\begin{abstract}
The shape of clouds has proven to be essential for classifying them. Our analysis of images from fair weather cumulus clouds reveals that, besides by turbulence they are driven by self-organized criticality (SOC). Our observations yield exponents that support the fact the clouds, when projected to two dimensions (2D), exhibit conformal symmetry compatible with $c=-2$ conformal field theory (CFT), in contrast to 2D turbulence which has $c=0$ CFT. By using a combination of the Navier-Stokes equation, diffusion equations and a coupled map lattice (CML) we successfully simulated cloud formation, and obtained the same exponents.
\end{abstract}

\maketitle

\section{Introduction}
The ``fractal'' or ``multi-fractal'' geometry as defined by Mandelbrot~\cite{Mandelbrot} is now well-established for clouds by many experimental and theoretical studies, and provides a powerful tool to classify them in terms of the circumstances in which they form. Self-affinity and scaling properties in clouds have been found from satellite images~\cite{Lovejoy}, and in particular in cumulus clouds~\cite{Austin} on several scales. Various observables were shown to exhibit scaling behavior, like the area-perimeter relation~\cite{Lovejoy,Chatterjee,Savigny,Austin,Malinowski,Batista,Madhushani}, the nearest neighbor spacing~\cite{Joseph}, the rainfall time series~\cite{Olsson}, cloud droplets~\cite{Malinowski2}, and the distribution function of geometrical quantities~\cite{Benner,Rodts,Yano,Gotoh}.
After these observations, and considering the multi-fractality of clouds~\cite{Lovejoy,Lovejoy2,Lovejoy3,Cahalan,Gabriel,Austin,Malinowski}, attempts for classifying clouds into universality classes were carried out based on cloud field statistics~\cite{Lovejoy4,Tessier,Pelletier} and cloud morphology~\cite{Sengupta}. The spatial distribution of clouds and their scale-invariant inhomogeneities on radiative fluxes and albedo~\cite{Killen,Sengupta,Aida,Claussen,Parol} has stimulated several theoretical models about radiative fluxes~\cite{Davies2,Welch,Kite}, often focusing on the scale-invariant radiation field from the clouds~\cite{Marshak}, and also on longwave irradiance and albedo in cumulus~\cite{Killen} and  stratocumulus~\cite{Cahalan3} clouds.\\
The models for clouds can be classified into three categories: turbulence-based models, cellular automata heuristic ones, and heuristic models. The models based on atmospheric fluid dynamics~\cite{Miyazaki,Wang,Yano2}, the scaling (or fractal geometry) in turbulent flows~\cite{Richardson,Hentschel,Malinowski}, large eddy simulations for cumulus convective transport~\cite{Siebesma,Zhao}, and stochastic models based on cascade processes~\cite{Schertzer,Tessier} belong to the first category. On the other hand, cellular automata~\cite{Nagel}, and computer graphics techniques for modeling clouds\cite{Gardner,Cianciolo,Bouthors} belong to the second category.
Other studies, like the application of the Kardar-Parisi-Zhang (KPZ) equation (inspired by the scale-invariant roughness of the top of the clouds)~\cite{Pelletier}, the diamond-square algorithm describing the fractal properties of cloud edges~\cite{Lohmann}, and also the midpoint displacement algorithm~\cite{Cai} are heuristic methods which can be incorporated into the third category. The estimated fractal dimensions emanating from these models are in fair agreement with the relative turbulent diffusion model~\cite{Rys}.\\
Despite of this huge literature, we are yet far from a complete understanding of cloud dynamics and its statistical properties. From the above mentioned observations, one concludes that clouds are characterized by scaling relations and exponents, which are robust under most circumstances, suggesting that clouds organize themselves into critical states without the need of tuning any parameter. Self-organized criticality (SOC) in the atmosphere and in clouds was first detected by Peters \textit{et al.} by analyzing precipitation~\cite{Peters}. Here we uncover the SOC state of clouds directly by analyzing earth-to-sky images of cumulus clouds under fair air conditions in Ardabil city, especially by analyzing the cloud boundaries, introducing a threshold to identify individual clouds on scales between tens of meters and several kilometers, i.e. the scale at which the turbulent scaling exponents change (as proposed by Beyer \textit{et. al.}~\cite{Beyer}), although other studies presented evidence for the absence of such a scale break~\cite{Lovejoy9}. The analysis of the level lines of two dimensional cloud fields strongly suggests that they are in an SOC state, which is also confirmed by a Schramm-Loewner evolution (SLE) analysis, finding that they belong to the sandpile universality class, i.e. $c=-2$ conformal field theory (CFT). We believe that in order to explain these observations on cumulus clouds, one needs to consider two ingredients at the same time: turbulence and SOC. We show that this can be achieved using the coupled map lattice (CML) which is an extended cellular automaton, based in our case on solving the Navier-Stokes (NS) equations along with some diffusion equations~\cite{Miyazaki}, which yields exponents consistent with the observed ones. 
\begin{figure*}[t]
	\centerline{\includegraphics[scale=0.48]{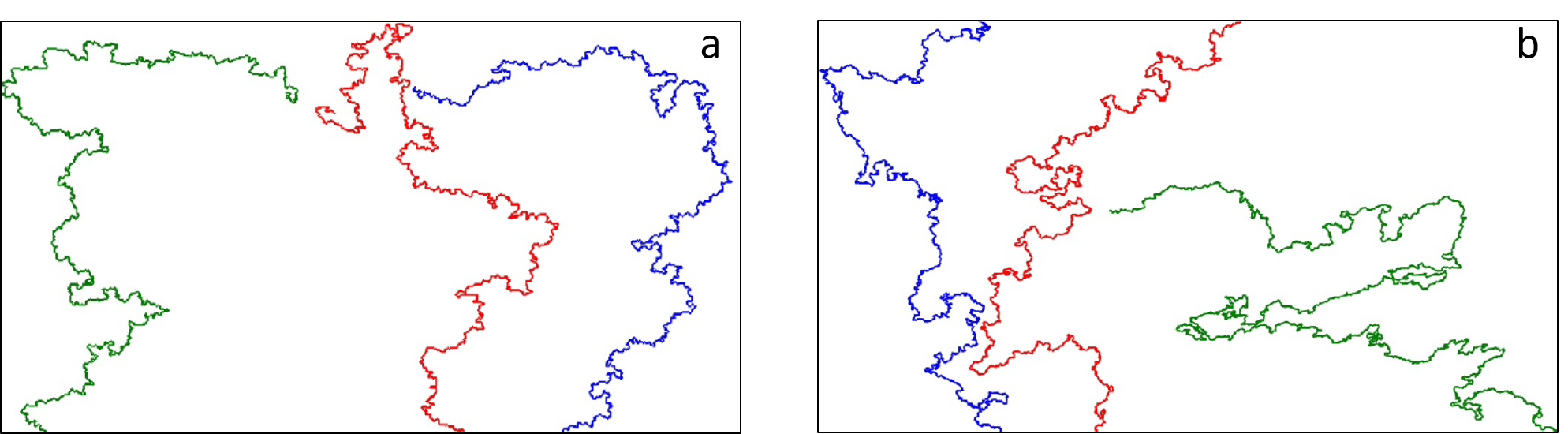}}
	\caption{ Three samples of (a) LERW  and (b) the boundaries of cumulus clouds.}
	\label{fig:Boundary_samples}
\end{figure*}
\section{Results}
\subsection{Observations with Visible Light}
During the daylight hours of Ardabil, we took (nearly vertical) images with a sky imager (a digital CMOS camera with a $18-140mm$ lens), realizing a resolution of $6000\times 4000$ pixels, under fair-air (small wind, no precipitation) conditions in June-July 2018. For more details and the air conditions at the time capturing the images see Table SI, and also Figs. S1, S2 and S3 of the supplementary material. We then converted the images to color triplets, so that each pixel has three numbers for the strength of red, green and blue, called RGB map. The images are converted to scalar fields (grey scale), low (high) values indicating clear (cloudy) pixels, and analyzed as correlated landscapes by focusing on the properties of their contour lines obtained at various thresholds. The contour lines corresponding to the lower thresholds are close to the physical boundary of the clouds, which is visible from the earth, see Figs. S2e and S2f in the supplementary material. Therefore, we interpret the contour line corresponding to a low enough threshold as the external perimeter of the cloud, i.e. the cloud boundary. We observed an abrupt change of intensity at cloud boundaries making their identification easy. As in any scalar landscape, for each photo there exists an intensity threshold at which the contour line percolates, i.e. becomes open, otherwise it is closed. Let $I_{bs}$ and $I_m>I_{bs}$ be the intensity of the blue sky and the maximum intensity inside the cloud respectively (generally higher intensities are from the pixels inside the clouds). The percolated contour lines arise approximately at the relative intensity $\frac{I_{th}}{I_m}=0.5$, which identifies contour lines close to the physical cloud boundaries (used in Figs. S2e and S2f in the supplementary material). Hereafter we call these contour lines the cloud \textit{boundary contour lines (BoCL)}, which are of crucial importance in this paper. Indeed, BoCL (which are open traces) are just the contour lines at the percolation threshold, which is well-known in correlated landscapes. Changing the relative intensity changes the contour line, but we observed that the variation is minimal in the interval $0.4\lesssim \frac{I_{th}}{I_m}\lesssim 0.6$. The contour lines that are out side of this range ($\frac{I_{th}}{I_m}\gtrsim 0.6$), called the \textit{bulk contour lines (BuCL)}, are closed traces having slightly different statistical properties, see Figs. S3a and S3b in the supplementary material. \\

To make the samples as independent as possible from each other, we took one photo every hour. The statistical properties of the BoCL resemble those of the traces of the loop-erased random walk (LERW)~\cite{Lawler}, see the samples depicted in Fig.~\ref{fig:Boundary_samples}. The projection to two dimensions enables us to use the power of CFT in 2D to characterize and classify the lines in universality classes.\\
We performed a fractal analysis for both BoCLs and BuCLs, obtaining the fractal dimension ($D_f$) in two ways: sandbox method and end-to-end distance statistics, see Fig. S4 in the supplementary material for details. The winding angle ($\theta$) statistics of the random traces as an important property of scale invariant loop-less paths~\cite{Duplantier,Wieland} is explored. We note that the BuCLs are basically closed traces over the whole sky, making them appropriate to be analyzed by the so called ``loop green function''~\cite{Kondev} (LGF) analysis. However, since the clouds span over various scales in the sky, in some rare cases the camera can cover only a part of them, i.e. the traces touch the image boundaries in these cases. In these situations, the trace can be interpreted as an open one, and the standard statistical analysis for open traces can be used~\cite{Najafi0}. The inverse is also true, i.e. although the BoCLs are often open (percolating) traces, there are some rare cases where the cloud is small, so that its corresponding BoCL is non-percolating. We consider only BoCLs that percolate. We apply the SLE mapping to BoCLs to numerically estimate the diffusivity parameter and equivalently the central charge of CFT which allows determining the CFT class. \\
\begin{figure*}[t]
	\centerline{\includegraphics[scale=0.5]{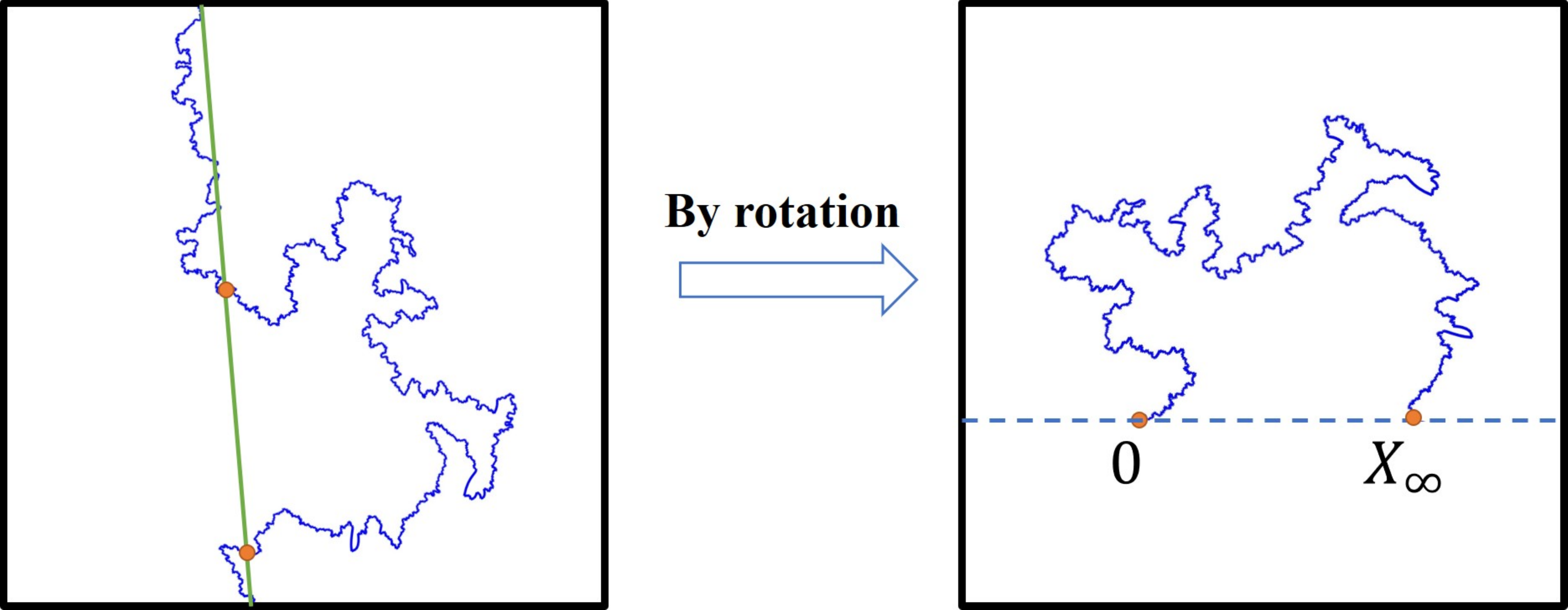}}
	\caption{The schematic representation of the process of generating a curve starting and ending on the horizontal (real) axis.}
	\label{fig:rotation}
\end{figure*}
For open traces (BoCLs) the fractal dimension ($D_f$) is defined via the scaling relation between the length $l$ and the box size $L$ ($l\sim L^{D_f}$) in the sandbox method, whereas for closed traces (BuCLs) we used the relation $\left\langle \log l\right\rangle=D_f\left\langle \log r\right\rangle  $ where $r$ is the gyration radius of the closed trace. The fractal dimension of the BoCLs and the BuCLs of the cloud field are shown in Figs.~\ref{fig:Df}a and \ref{fig:Df}b respectively, where $R$ is end-to-end distance, $N$ is the number of steps along the trace, and $\bar{R}\equiv\sqrt{\left\langle R^2\right\rangle }\propto N^{\nu}$. The analysis for BuCLs is presented in Fig.~\ref{fig:Df}b, involving the fractal dimension and the distribution functions. Our analysis shows that $D_f^{\text{bd}}=1.248\pm 0.006$ (for the BoCLs) and $D_f^{\text{bk}}=1.22\pm 0.02$ (for the BuCLs). This $D_f^{\text{bd}}$ is compatible with the numerical estimation of the end-to-end distance exponent $\nu^{bd}$ of BoCLs which is $0.81\pm 0.01$ ($= \frac{1}{D_f^{\text{bd}}}$). Interestingly the fractal dimensions of the boundaries of two-dimensional images of the clouds are very close to the fractal dimension of LERW traces with $D_f=\frac{5}{4}$ and $\nu=0.8$~\cite{Lawler}. The LGF ($G(r)$) which is defined as the probability that two points at a distance $r$ belong to the same closed trace~\cite{Kondev}, shows logarithmic behavior with $r$, as seen in the lower inset of Fig.~\ref{fig:Df}a, as it is also the case for the LERW~\cite{Bauer1,Dhar}. \\
\begin{figure*}
	\centerline{\includegraphics[scale=0.8]{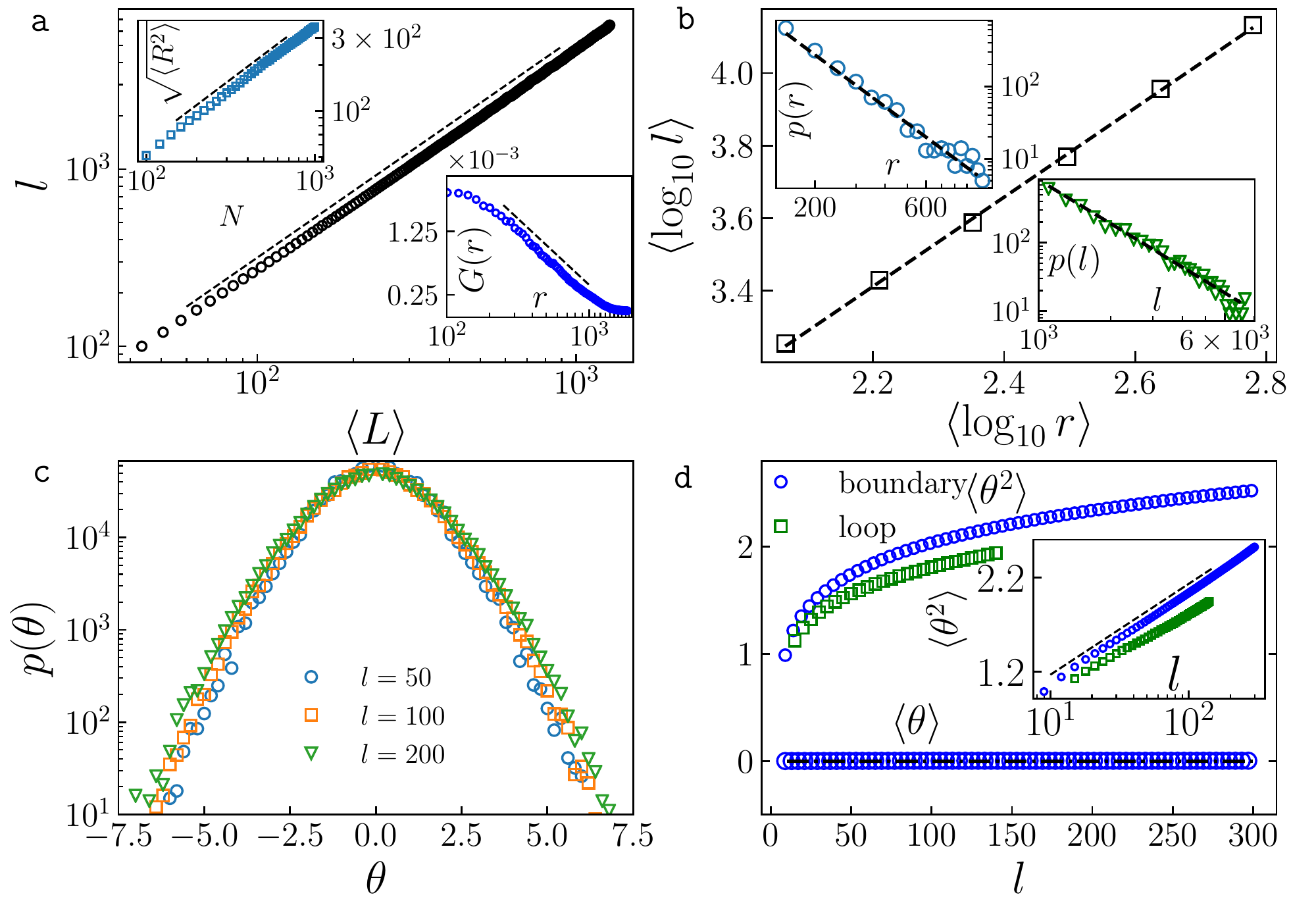}}
	\caption{The fractal properties of clouds from observational data. (a) Log-log plot of trace lengths $l$ in terms of $L$ (the box linear size) for BoCLs. The dashed line is a linear fit with slope $D_f=1.248\pm 0.006$. Upper inset is the end-to-end distance $\bar{R}$ in terms of $N$, and the lower inset is a semi-log plot of the LGF in terms of $r$, with the exponent $\nu=0.81\pm 0.01$. (b) Ensemble average of $\log l$ in terms of $\log r$ ($\left\langle \right\rangle $ means the ensemble average) with slope $D_f=1.22\pm 0.02$ for BuCLs. The log-log plot of the distribution functions of $r$ and $l$ are shown in the upper and lower insets, with exponents $\tau_r=2.12\pm 0.03$ and $\tau_l=2.38\pm 0.02$ respectively. (c) Semi-log plot of the distribution of the winding angle $p(\theta)$ in terms of $\theta$ for three BoCLs. (d) $\left\langle \theta^2\right\rangle $ and $\left\langle \theta\right\rangle $ for both BoCLs and BuCLs. The inset shows the semi-log plot of the variance in terms of $l$ for both cases with slopes $0.42\pm 0.01$ and $0.36\pm 0.02$ for boundary and bulk curves respectively.}
	\label{fig:Df}
\end{figure*}
The statistics of the winding angle $\theta$ is an important (in many cases the most precise) way to characterize fractal random curves. For BoCLs which are open traces we can use the following relation for the variance of $\theta$~\cite{Duplantier,Wieland}:
\begin{equation}
\text{Var}[\theta]=C +2\frac{D_f-1}{D_f}\ln l
\label{Eq:WA}
\end{equation}
where $\text{Var}[\theta]$ is the variance of $\theta$, and $C$ is an irrelevant constant. In the set up shown in Fig. S4a of the supplementary material, $\theta$ is the angle between the straight line between two points (at distance $l$) on the curve and the local slope of the trace. In this set up we fix a point (the point Q in the figure) and let P run over the curve, then $\theta$ is the angle between the local slope and the shown line. Figure~\ref{fig:Df}c shows the distribution function of $\theta$ (which is Gaussian), and Fig.~\ref{fig:Df}d shows $\left\langle \theta^2\right\rangle $ and $\left\langle \theta\right\rangle $ in terms of $l$ for the BoCLs (blue circles), revealing that the variances depend logarithmically on the size $L$ with the prefactor $0.42\pm 0.01$, to be compared with the prefactor for the LERW traces which is $0.4$. We applied this analysis to BuCLs by cutting the loop horizontally (at the middle) to have an open trace and apply the algorithm to a part (one-fourth) of the resulting trace. The green squares in Fig.~\ref{fig:Df}d show that the variance grows logarithmically with $L$ having a prefactor $0.36\pm 0.02$, which is compatible with the fractal dimension, i.e. $2\frac{D_f^{\text{bk}}-1}{D_f^{\text{bk}}}=0.36\pm 0.03$.\\

The SLE theory describes the critical behavior of 2D loop-less paths~\cite{Bauer1}. These non-intersecting curves which reflect the status of the system in question are supposed to have two properties: conformal invariance and the domain Markov property. Thanks to a deep connection between SLE and CFT, less-known conformally invariant models can be classified into CFT universality classes by applying the SLE mappings to interfaces of the model. For a good introductory review see references~\cite{Rohde1,Cardy}. One defines the \textit{uniformizing conformal maps} ($g_t(z)$) which are solutions of the stochastic Loewner's equation
\begin{equation}
\partial_{t}g_{t}(z)=\frac{2}{g_{t}(z)-\zeta_{t}},
\label{Loewner}
\end{equation}
\begin{figure*}
	\centerline{\includegraphics[scale=0.7]{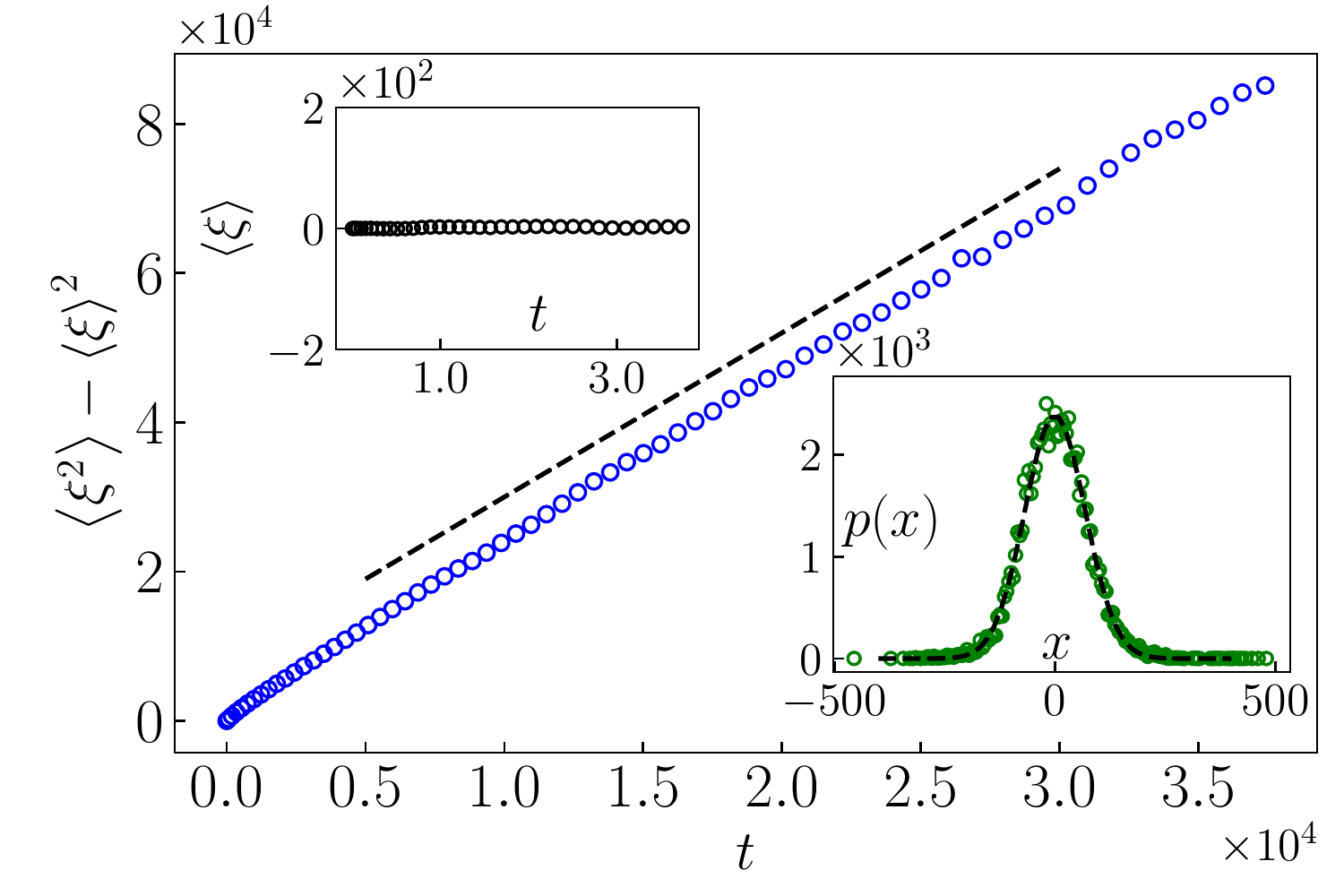}}
	\caption{The results for the SLE analysis for the BoCLs. The main panel shows the variance of the driving function in terms of ``time'', with the slope $\kappa=2.2\pm 0.2$. The upper inset shows that the mean value of the driving function is almost zero, and the lower one shows the distribution function of $\frac{\zeta_t}{\sqrt{2}}$ at time $t=10^4$, with a Gaussian fit corresponding to that time.}
	\label{fig:SLE}
\end{figure*}
in which the initial condition is $g_{t}(z)=z$  and the driving function $\zeta_{t}$ is proportional to a one dimensional Brownian motion $B_{t}$ i.e. $\zeta_{t}=\sqrt{\kappa}B_{t}$ in which $\kappa$ is called the diffusivity parameter, which is the quantity that classifies the conformally invariant models~\cite{Cardy}. The name ``\textit{uniformizing map}'' comes from the fact that $g_t$ uniformizes the conformally invariant trace (defined in the upper half plane) onto the real axis. We have used the algorithm invented by Bernard \textit{et al.}~\cite{Bernard2}, which is appropriate for conformal traces that hit the boundaries of the system (here the boundaries of the images). According to this algorithm one discretizes the random trace (therefore the driving function becomes piecewise constant), and sends the end point of the curve ($x_{\infty}$ at which the trace hits the real axis) to infinity using a simple M{\"o}bius map $\phi(z)=\frac{x_{\infty}z}{z-x_{\infty}}$, where $z=x+iy$ is the complex upper half plane. For details of the numerical application of the SLE theory, see the supplementary material (the discretized map is Eq. 4). For the traces that start and end on different boundaries, we first draw a straight line which connects the first and last point of the trace, and rotate it appropriately (by an angle given by this line, and interpreting the space above this line as the upper half plane) to have a trace which starts and ends on the real axis. See Fig.~\ref{fig:rotation} where the situation is shown schematically. In such situations, we first identify the straight line connecting the start and end point of the trace, and then cut the trace at the crossing points. The required trace is then obtained simply by rotating the system, so that the straight line becomes horizontal. By interpreting the space above this line as the upper half plane, we can use the above-mentioned algorithm. Obviously the rotational invariance, and also the reflection symmetry of the system allows us to consider and analyze all traces resulting from one cut (noting that each cut gives us more than one curve).
Fig.~\ref{fig:SLE} shows the variance of the driving function for BoCLs $\left\langle \zeta^2\right\rangle -\left\langle \zeta\right\rangle^2 $ which grows like $\kappa t$, with $\kappa=2.2\pm 0.2$, and $\left\langle \zeta\right\rangle=0 $ as required by conformal symmetry~\cite{Bauer1}. The fact that the diffusivity parameter for LERW traces is $\kappa=2$~\cite{Najafi2}, confirms the above-mentioned hypothesis that the traces of BoCLs have the same symmetry as the LERW traces. We note that the corresponding fractal dimension calculated by the relation $D_f=1+\frac{\kappa}{8}$~\cite{Cardy} is $1.27\pm 0.02$, consistent with the values we obtained before. The exponents are shown in table~\ref{tab:exponents}.\\

We thus have strong numerical evidence that the external perimeter of 2D projections of cumulus clouds might belong to the universality class of LERW traces, which does have a relation to SOC. Indeed it has been argued that the external frontiers of sandpiles (as a prototype example of SOC systems) are LERW traces~\cite{Majumdar,Mahieu}, both of them having the same central charge $c=-2$ CFT. Taking into account the recent finding that the boundaries of two-dimensional projections (like the images here) of three dimensional sandpiles also seem to belong to the universality class of LERWs~\cite{Najafi1}, we suggest that, SOC is an important ingredient of the evolution of the cumulus clouds. We confirm this claim by simulating the system, which is the aim of the next section.\\

\begin{table}
	\center
	\caption{Numerical values of the diffusivity parameter $\kappa$ and the fractal dimension $D_f$ obtained by various methods. The values for the ordinary 2D BTW model are also shown for comparison.}
	\label{tab:conditions}
	\begin{tabular}{|c | c | c | c |}
		\hline method & $\kappa$ & $D_f$ & $2(D_f-1)/D_f$ \\
		\hline Sandbox & $2.0\pm 0.2$ & $1.248\pm 0.006$ & $0.4 \pm 0.01$ \\
		\hline Winding angle & $2.1 \pm 0.2$ & $1.266 \pm 0.02$ & $0.42 \pm 0.01$ \\
		\hline SLE (IM) & $2.2 \pm 0.2$ & $1.27 \pm 0.02$ & $0.43 \pm 0.02$ \\
		\hline 2D BTW & $2$ & $1.25$ & $0.4$ \\
		\hline
	\end{tabular}
\label{tab:exponents}
\end{table}

\subsection{Simulation Results}
Here we model the evolution of cumulus clouds. These clouds result from atmospheric convection and form as the air warms up at the surface and begins to rise, resulting in a decrease of temperature and a rise of humidity. At a threshold, called lowest condensation level (LCL), at which the relative humidity reaches $100\%$, the condensation to the wet phase starts, and the released latent heat warms up further the surrounding air parcels, resulting in convection. The simulation of this complex dynamics requires a substantial reduction of the degrees of freedom to become feasible to be carried out on a computer. This reduction however should maintain the key phenomena like turbulence and SOC. In a coupled map lattice (CML), as an extension of a cellular automaton~\cite{Miyazaki,Kaneko}, one considers several state variables in space, discretized on a lattice. This model is based on the update of the state of ``particles'' on the lattice, which are the water vapor particles in the clouds. To be more precise, we discretize form a lattice, each cell of which contains a huge number of water vapor particles, described mathematically by the vapor density in it, having a single (drift) collective velocity. Avalanches, as the typical ingredients of SOC dynamics result from the accumulation of particles on a lattice site, and subsequent toppling (particle release) beyond a threshold (the saturation humidity). This is implicitly realized in the CML model where, beyond this threshold, the accumulated moist particles transit from the vapor phase into the liquid phase (and vice versa), reducing the number of particles in the vapor phase. The particles in the liquid-phase then diffuse to the neighboring sites due to the rise in density. Here the saturation humidity plays the role of the threshold beyond which the particles start to move to neighboring sites. Let us consider the NS equation
\begin{equation}
\left( \frac{\partial}{\partial t}+\textbf{v}.\nabla\right)\textbf{v}=-\frac{1}{\rho}\nabla P+\eta \nabla^2 \textbf{v}
\label{Eq:NS}
\end{equation}
\begin{figure*}[t]
	\centerline{\includegraphics[scale=0.45]{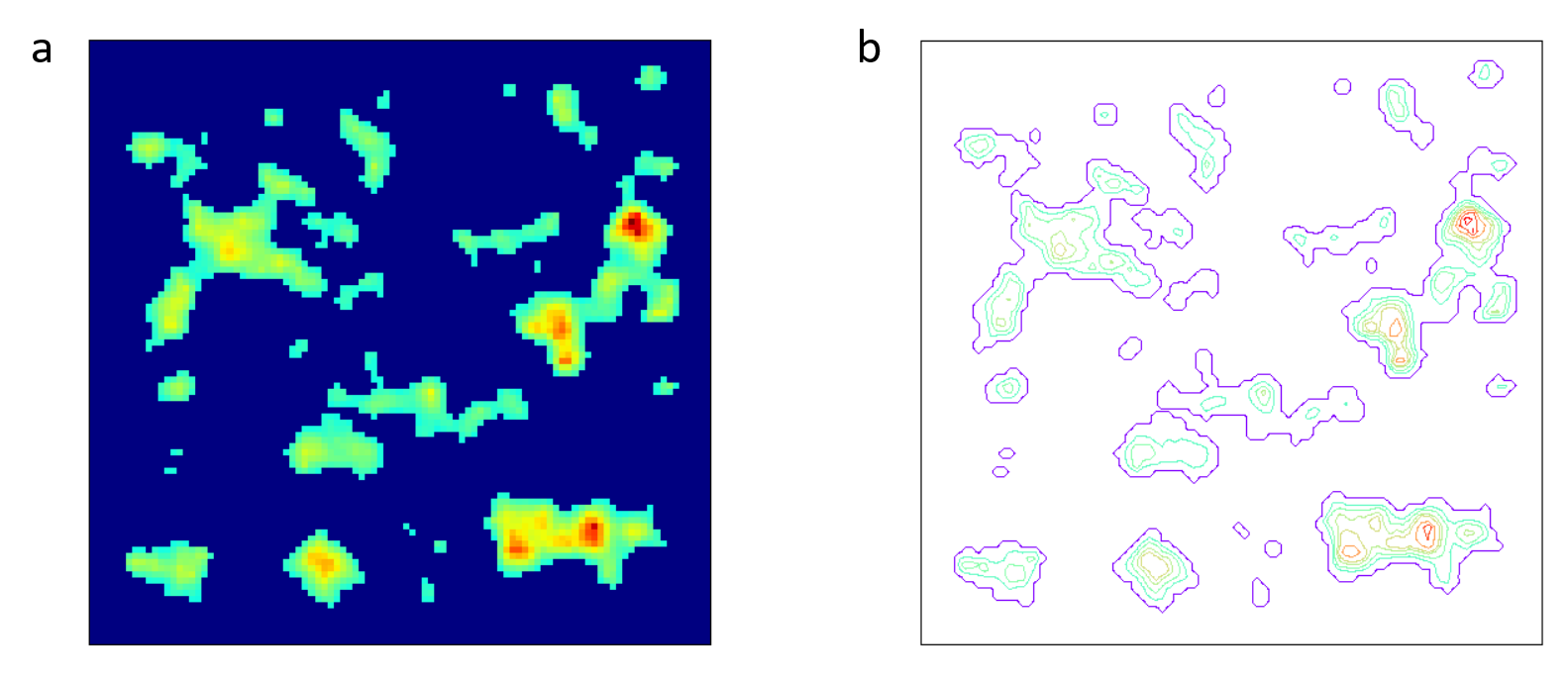}}
	\caption{(a) 2D image of cloud simulation and (b) its contours.}
	\label{fig:simulation_sample}
\end{figure*}
where $\textbf{v}$ is the velocity field, $P$ the pressure, $\eta$ the viscosity, and $\rho$ the density. Along with this equation, we use the incompressibility of the atmosphere, which implies $\nabla.\textbf{v}=0$ (note that, actually the atmosphere is a compressible fluid, however for velocities below the speed of sound the density changes are small enough to consider it as incompressible). In our CML we consider a $L_x\times L_y\times L_z$ lattice, for which we attribute to each site the density of water vapor $\rho_w(x,y,z)$ and liquid water $\rho_l(x,y,z)$, the pressure $p(x,y,z)$, the velocity field $\textbf{v}(x,y,z)$, and the temperature $T(x,y,z)$ which are updated depending on the variables on the adjacent lattice sites like in cellular automata. Initially a bubble of air is heated from below, which leads to its rise upwards, i.e. in $z$ direction. At each update, we consider four effects: 1- the velocity update due to Eq.~\ref{Eq:NS}, 2- the convection, for which the fields are updated due to the fluid movement, 3- the diffusion of water vapor, 4- the phase transition, according to which, when the water vapor of a site exceeds the saturation threshold, it condensates to the liquid phase. Note that the updates are made in parallel.\\
Instead of solving equations listed in Eq.~\ref{Eq:NS}, in the CML method one considers some other approximate models~\cite{Miyazaki,Kaneko}. To get rid of the difficulties of handling the equation $\nabla .\textbf{v}=0$, in the CML method one uses $\nabla \left( \nabla .\textbf{v}\right) $, which replaces the gradient of pressure by a phenomenological proportionality constant $k_p$, called coefficient of pressure~\cite{Miyazaki,Kaneko}. This avoids the complications of modeling the pressure, the mass conservation equation and the incompressibility. Denoting $V_i(x,y,z)\equiv V_i(x,y,z;t)$ ( $i$th component the velocity $\vec{V}$) and $\tilde{V}_i(x,y,z)\equiv V_i(x,y,z;t+1)$ (the velocity at the next time step), the update of velocity is
\begin{equation}~\label{Eq:laplas}
\begin{split}
\tilde{V}_i(x,y,z) =& V_i(x,y,z) + k_v \nabla^2 V_i(x,y,z) \\
&+k_p \nabla(\nabla.\vec{V})_i
\end{split}
\end{equation}
where $k_v$ is the viscosity ratio, and $k_p$ is the coefficient of the pressure effect. In reality $k_v$ and $k_p$ (having dimension of $L^2$) are related to the environmental conditions, such as pressure and temperature, and determine the LCL. In simulations they are used to tune the LCL point, and according to our observations, they do not have any decisive role in the pattern of clouds after LCL. In the simulations we consider $k_v= 5.73\times10^{-2}$ and $k_p=5.73\times10^{-2}$~\cite{Yanagita,Yanagita2}, for which the LCL forms at $z_{LCL}\approx 50 $. Using the discretized Laplacian and for $\nabla(\nabla.\vec{V})_i$ we update the velocities~\cite{Miyazaki,Kaneko}, see supplementary material for more information. The physics of advection (as the second step) is implemented by transporting the state values at the lattice point $(x,y,z)$ to a new position $(x +V_x,y +V_y,z+V_z ) =(i +\delta x , j + \delta y , k +\delta z ) $, where $i$, $j$ and $k$ are the integer part and $\delta_x$,  $\delta_y$ and $\delta_z$ are the fractional portions. The variables are updated according to the ``weights'' recieved from adjacent sites in the previous step. These weights are actually the fraction of particles that are leaving a site. We notice that for the stability of the solution, we should have  $V_x,V_y,V_z <1.0$, otherwise we should use Stam’s method \cite{Blanz} to make it stable. We observed that our simulation is stable and does not require this stabilization. In the third part we consider the diffusion of water vapor $\rho_w(x,y,z)$ which is implemented by using a discrete version of the diffusion equation
\begin{equation}
\tilde{\rho}_w(x,y,z) = \rho_w(x,y,z) + k_w \nabla^2 \rho_w(x,y,z),
\end{equation}
\begin{figure*}[t]
	\centerline{\includegraphics[scale=0.76]{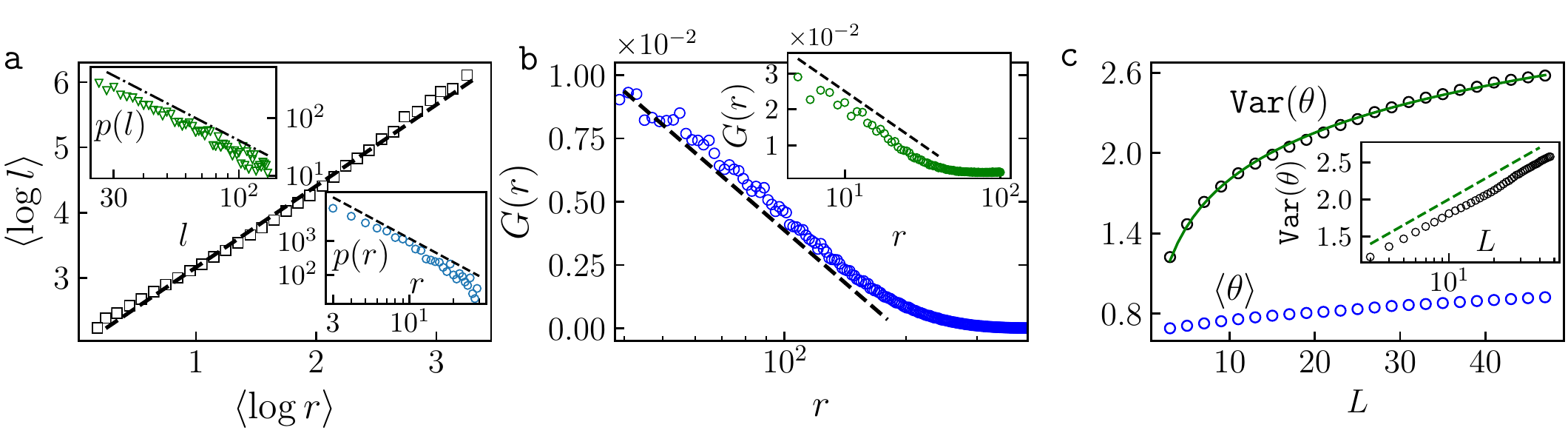}}
	\caption{The results of simulations of clouds using the CML method. (a) $\left\langle \log l\right\rangle $ in terms of $\left\langle \log r\right\rangle $ with the slope $D_f=1.247\pm 0.016$. The upper and lower insets show the log-log plot of the distribution functions of  $l$ and $r$ with exponents $\tau_l=2.14\pm 0.05$ and $\tau_r=2.35\pm 0.06$ respectively. (b) Semi-log plot of the LGF in terms of $r$. (c) Variance of $\theta$ and its average, where the inset shows the semi-logarithmic graph with slope $0.506\pm 0.015$.}
	\label{fig:PrnG}
\end{figure*}
where $ k_w$ is the diffusion coefficient for water vapor and $ \nabla^2 \rho_w(x,y,z) $ is the Laplacian of the density that is calculated in the same way as Eq.(\ref{Eq:laplas}). Just like $k_v$ and $k_p$, $k_w$ (as well as $\alpha$ and $\beta$ to be defined in the following) affect only the LCL, and we set it $ k_w = 0.3$ following Refs.~\cite{Cussler,Yanagita,Yanagita2}.\\
As a last part of simulation we implement the physics of the phase transition. For this end let us consider $\rho_{\text{max}}$ as the maximum amount of water vapor density that can exist at temperature $T$~\cite{Miyazaki} which is estimated to be
\begin{equation}
\rho_{\text{max}}(T) =  \frac{217.0e^{[19.482 - \frac{4303.4}{T-29.5}]}}{T}. 
\end{equation}
The density of water vapor $\tilde{\rho}_w (x,y,z)$, the water liquid $\tilde{\rho}_l (x,y,z)$ and the temperature $\tilde{T}(x,y,z)$ at the next step change due to the phase transition as follows
\begin{align}
&\tilde{\rho}_l = {\rho}_l +\alpha({\rho}_l-{\rho}_{\text{max}}), \\& \tilde{\rho}_w = {\rho}_w -\alpha({\rho}_w-{\rho}_{\text{max}}), \\& 
\tilde{T} = {T} -\beta({\rho}_w-{\rho}_{\text{max}})
\end{align}
where $\alpha$ and $\beta$ are the phase transition rate and adiabatic expansion rate respectively, which are set to $\alpha=\beta=10^{-2}$~\cite{Yanagita,Yanagita2}. Some simulated samples are shown in Fig.~\ref{fig:simulation_sample}.\\
We set $\rho_l (x,y,z;t=0) =0$ where the air parcels start to rise from the bottom of the lattice. We start the simulation at $t=0$ by setting random values for $\rho_w (x,y,z;t=0)$ on the $x-y$ plane, and also setting the velocities $V_x(x,y,z;t=0)$ and $V_y(x,y,z;t=0)$ randomly in the range between $-1$ and $1$ for each site that has $\rho_w (x,y,z;t=0)\neq 0$. This choice assures that the particles move symmetrically to the right and left. To assure that the particles ascend in $z$-direction, we consider at the beginning $V_z(x,y,z;t=0)$ randomly and uncorrelated in the range $(0,1)$. We set the temperature to $300K$ in the $x-y$ plane ($z=0$) and decrease it linearly in $z$-direction. We considered $L_x=L_y=150$ and $L_z=70$, and start with a random density of water vapor on the bottom, with (uncorrelated) random velocities in $x$ and $y$ directions. Then we let the system evolve forming cumulus clouds. The ensemble averages have been taken over $10^4$ samples. We fix the temperature at $300K$ on the $x-y$ plane at $z=0$, and decrease it linearly with increasing altitude, as an acceptable approximation for the known temperature profile~\cite{Miyazaki}.\\ 
In Fig.~\ref{fig:PrnG}a we show the results for the fractal dimension and the distribution functions for the gyration radius $r$ and the loop length $l$ for two-dimensional projections. We find the fractal dimension $D_f=1.247\pm 0.016$, and also $\tau_l=2.14\pm 0.05$ and $\tau_r=2.35\pm 0.06$ which are in excellent agreement with the observed exponents. Also the Green's function is found to be logarithmic (Fig.~\ref{fig:PrnG}b). We analyzed the winding angle statistics of the boundaries of 2D projections, presented in Fig.~\ref{fig:PrnG}c. Here $\theta$ is the local change of slope, and the variance is taken over a box of linear length $L$. Using the fact that $D_f=1+\frac{\kappa}{8}$ and $l\sim L^{D_f}$, and also Eq.~\ref{Eq:WA} one can easily show that $Var[\theta]=c'+\frac{\kappa}{4}\ln L$, where $c'$ is an unimportant constant~\cite{Daryaei}. By a linear least-square fitting of the semi-logarithmic plot in Fig.~\ref{fig:PrnG}c we found that the corresponding diffusivity parameter is $\kappa=2.024\pm 0.06$. We see that, the exponents that we extracted from the CML model are in agreement with the observed exponents.\\
Summarizing we showed that the two-dimensional projection of cumulus clouds forms fractal borders with a fractal dimension consistent with $5/4$ and fulfills the conditions of the SLE mapping and the winding angle, pointing towards a CFT with central charge $c = -2$ which would be in the universality class of the loop-erased random walk. Interestingly this is not the universality class of level sets in 2D turbulence, for which the central charge is $c = 0$. We formulated a model for cumulus clouds based on fluid dynamics and SOC which could reproduce the results obtained from the image analysis. In the future it would be interesting to also verify with our model other properties of cumulus clouds, like their three-dimensional shape, their spatio-temporal evolution and their size distribution, which will however require substantially more computer time.

\captionsetup[figure]{name=FIG.}
\captionsetup[table]{name=TABLE}
\renewcommand\thefigure{S\arabic{figure}}
\renewcommand\thetable{S\arabic{table}}

\definecolor{orange}{rgb}{1,0.5,0}
\definecolor{brown}{rgb}{0.65, 0.16, 0.16}
\definecolor{phlox}{rgb}{0.87, 0.0, 1.0}
\newpage
\textbf{SUPPLEMENTARY MATERIAL for\\
	Self-organized criticality in Cumulus Clouds}\\

In this text we present some details of the observation and simulation of cumulus clouds. 
\begin{figure*}[t]
	\centerline{\includegraphics[scale=.13]{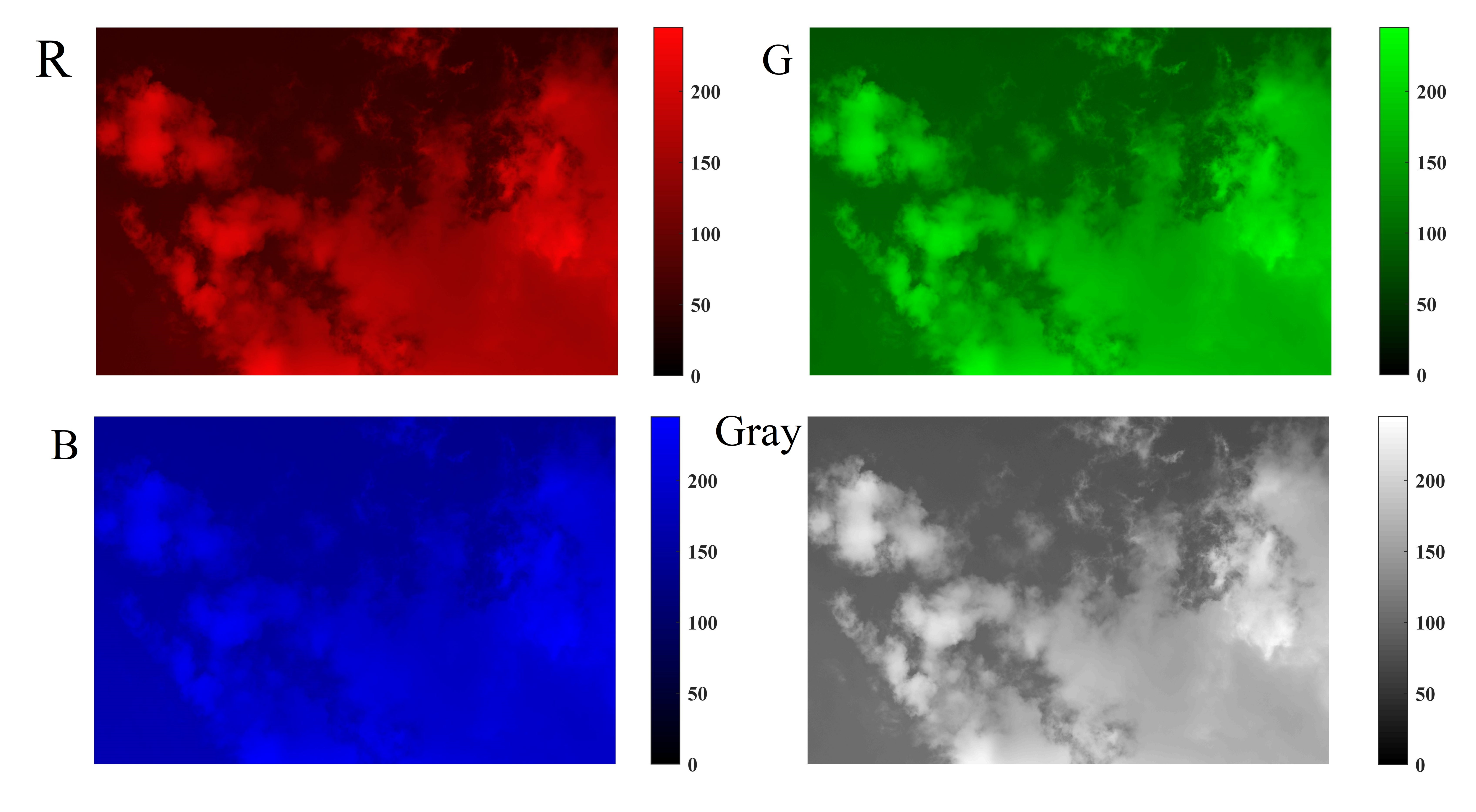}}
	\caption{RGB  and gray maps of an image that was taken from cumulus clouds. the color bars show the intensity of light of each color}
	\label{fig:RGB}
\end{figure*}
\section{Statistical analysis of the observational data}
\begin{figure*}[t]
	\centering
	\begin{subfigure}{0.49\textwidth}\includegraphics[width=\textwidth]{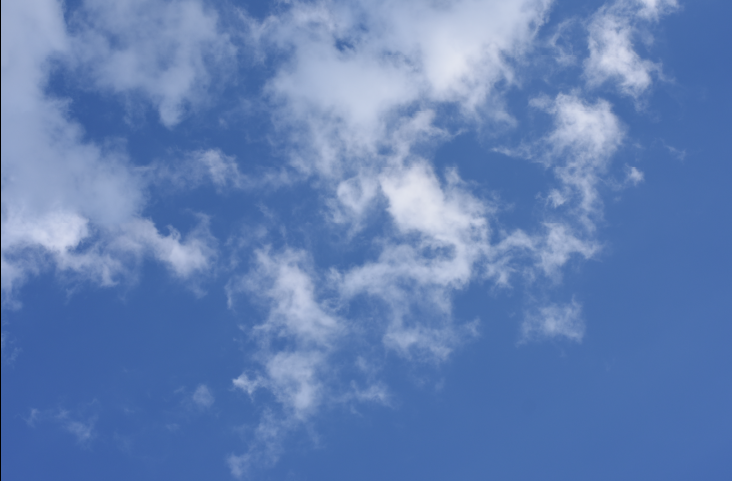}
		\caption{}
		\label{fig:Sample0}
	\end{subfigure}
	\begin{subfigure}{0.49\textwidth}\includegraphics[width=\textwidth]{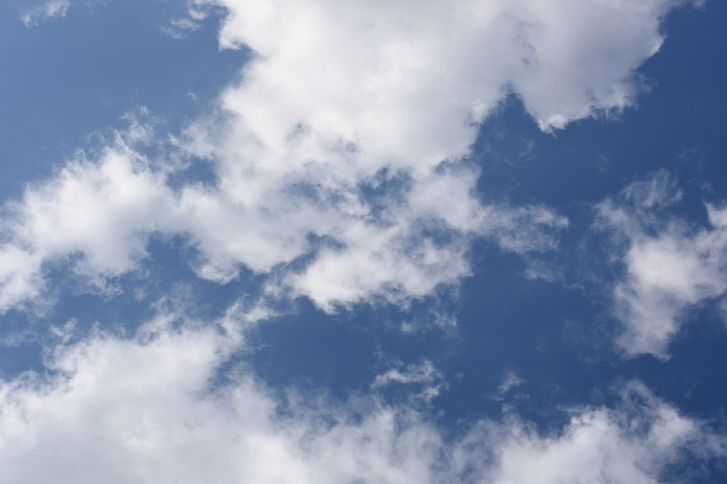}
		\caption{}
		\label{fig:Sample1}
	\end{subfigure}
	\begin{subfigure}{0.49\textwidth}\includegraphics[width=\textwidth]{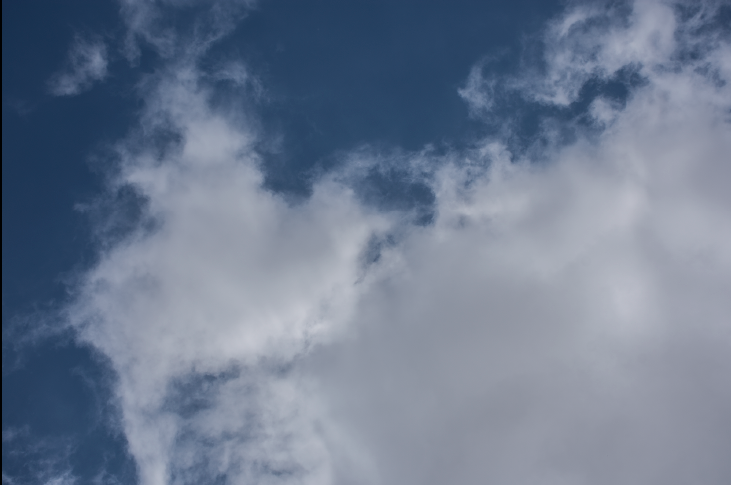}
		\caption{}
		\label{fig:Sample2}
	\end{subfigure}
	\begin{subfigure}{0.49\textwidth}\includegraphics[width=\textwidth]{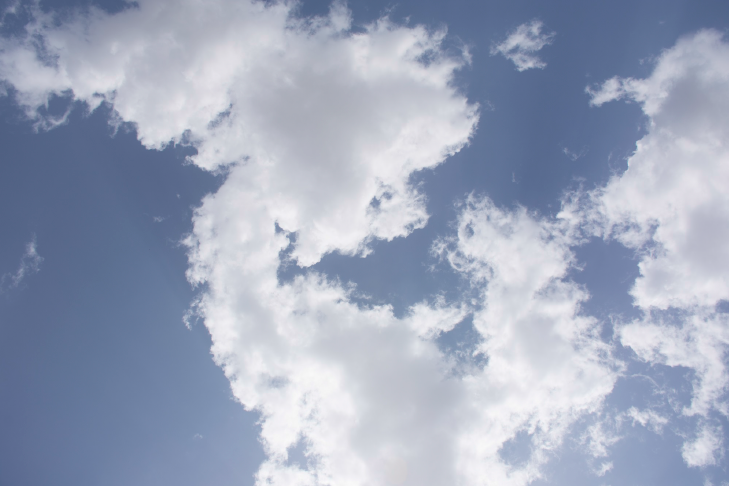}
		\caption{}
		\label{fig:Sample3}
	\end{subfigure}
	\begin{subfigure}{0.49\textwidth}\includegraphics[width=\textwidth]{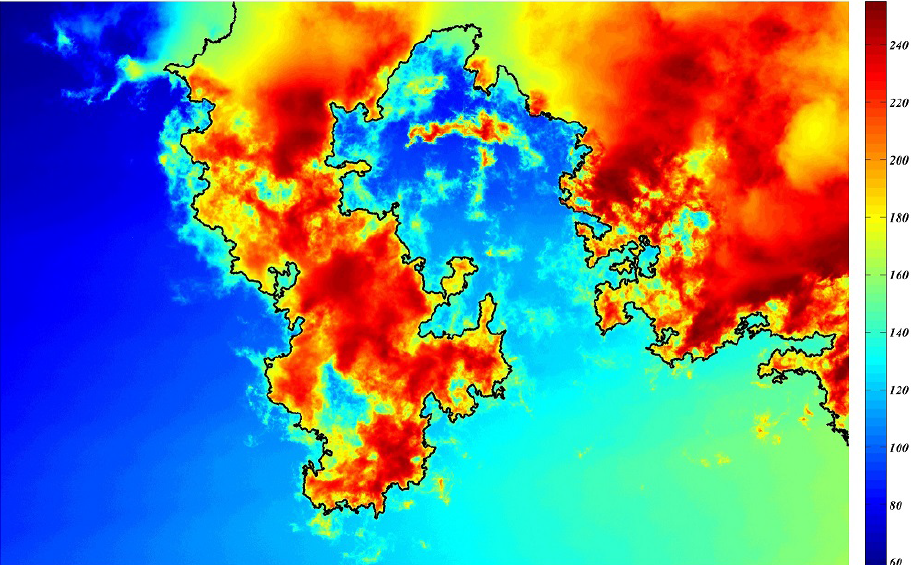}
		\caption{}
		\label{fig:Sample4}
	\end{subfigure}
	\begin{subfigure}{0.49\textwidth}\includegraphics[width=\textwidth]{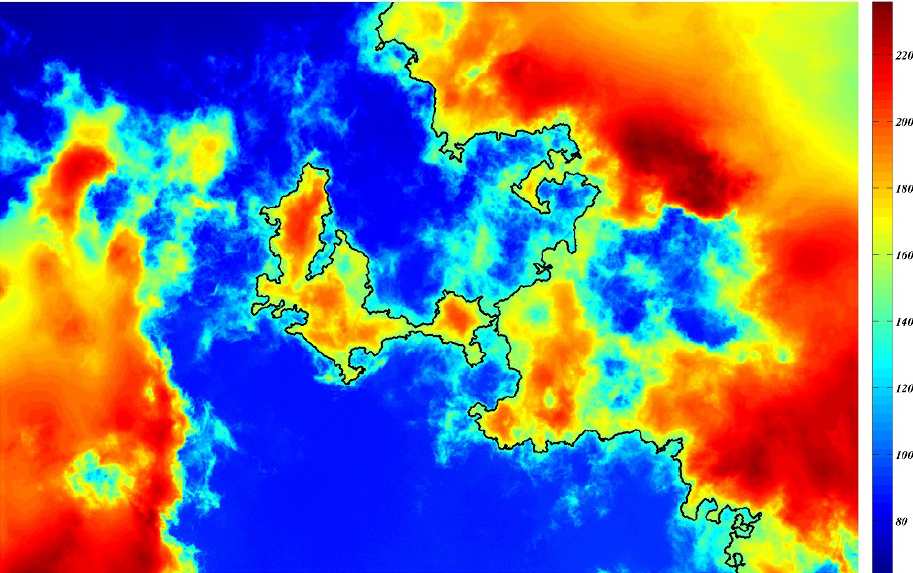}
		\caption{}
		\label{fig:Sample5}
	\end{subfigure}
	
	\caption{(Color online) (a) (b) (c) and (d) are four sample pictures from cumulus clouds. (e) and (f) the boundaries of two samples of cumulus clouds in gray map.}
	\label{fig:samples1}
\end{figure*}

\begin{figure*}[t]
	\centering		
	\begin{subfigure}{0.49\textwidth}\includegraphics[width=\textwidth]{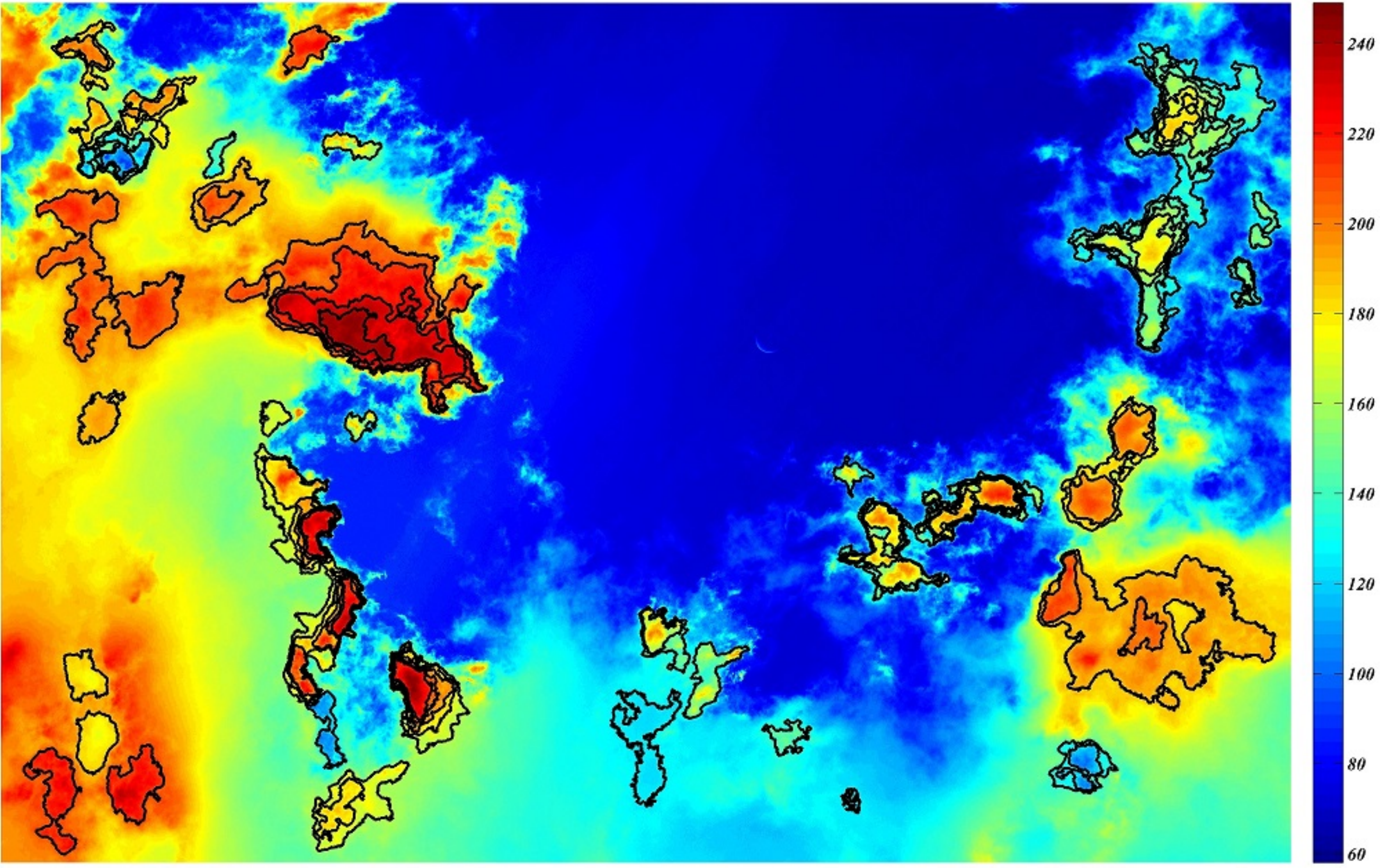}
		\caption{}
		\label{fig:Sample7}
	\end{subfigure}
	\begin{subfigure}{0.49\textwidth}\includegraphics[width=\textwidth]{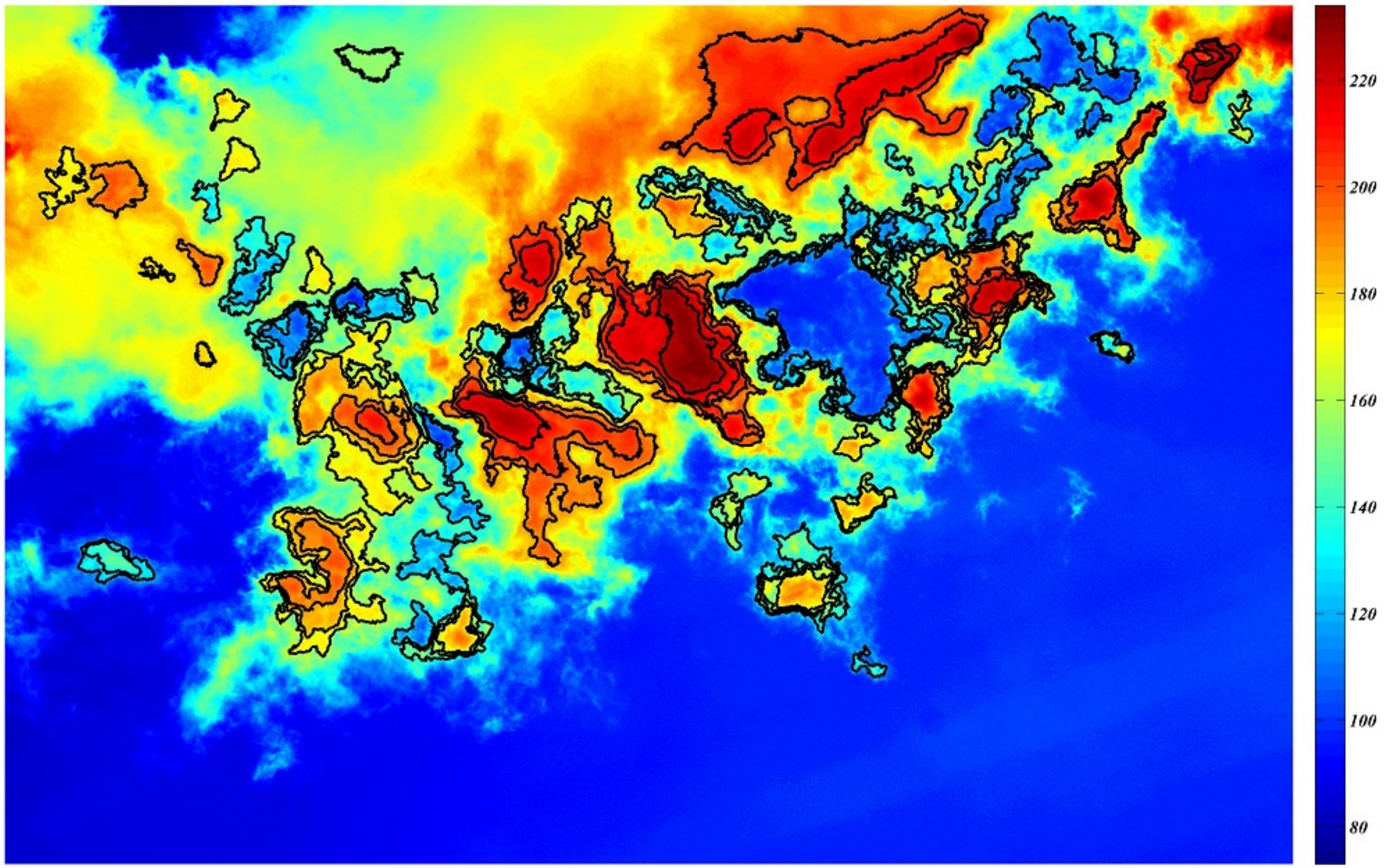}
		\caption{}
		\label{fig:Sample8}
	\end{subfigure}
	\caption{(Color online) (a) and (b) in contour lines inside the cumulus clouds.}
	\label{fig:samples2}
\end{figure*}

The weather conditions under which the photos were taken have been gathered in Table~\ref{tab:conditions1}. Some photos focused only on a part of a cloud, since the clouds were very large spanning most part of the sky. Each photo consists of three matrices showing the intensity of red (R matrix), green (G matrix) and blue (B matrix), called RGB map. To simplify our analysis we use a gray map instead of the RGB map i.e. the mean of the RGB, see Fig.~\ref{fig:RGB}. Figures~\ref{fig:samples1}a-d show some samples, and two samples of boundary contour lines (BoCL) are shown in Figs.~\ref{fig:Sample4} and~\ref{fig:Sample5}. The bulk contour lines (BuCL) are shown in Figs.~\ref{fig:Sample7} and~\ref{fig:Sample8}, which are closed traces.

\begin{table*}[t]
	\begin{tabular}{c | c | c | c | c | c | c | c}
		\hline date (2018) & temperature & Dew point & humidity & pressure & wind speed & precipitation & $\alpha$ (deg)\\
		\hline $6$th June & $68-82^{\circ F}$ & $43-52^{\circ F}$ & $25-46 \%$ & $25.7$ in & $7-16$ mph (WNW) & $0.0 $ in & 15:50-38:20 \\
		\hline $9$th June & $63-66^{\circ F}$ & $54-57^{\circ F}$ & $68-73 \%$ & $25.6$ in & $12-18$ mph (ENE) & $0.0$ in & 15:40-38:00 \\
		\hline $10$th June & $61-70^{\circ F}$ & $48-52^{\circ F}$ & $50-72 \%$ & $25.5$ in & $5-20$ mph (E) & $0.0 $ in & 15:40-38:00 \\
		\hline $14$th June & $64-73^{\circ F}$ & $50-54^{\circ F}$ & $50-68 \%$ & $25.6$ in & $9-16$ mph (ENE) & $0.0 $ in & 15:30-37:40 \\
		\hline $17$th June & $66-77^{\circ F}$ & $50-54^{\circ F}$ & $41-60 \%$ & $25.5$ in & $5-7$ mph (WNW) & $0.0 $ in & 15:20-37:30 \\
		\hline $18$th June & $70-77^{\circ F}$ & $52-57^{\circ F}$ & $44-53 \%$ & $25.5$ in & $12-14$ mph (E) & $0.0$ in & 15:20-37:30 \\
		\hline $21$th June & $59-72^{\circ F}$ & $41-43^{\circ F}$ & $33-51 \%$ & $25.7$ in & $7-18$ mph (NE) & $0.0$ in & 15:20-37:20 \\
		\hline $23$th June & $68-75^{\circ F}$ & $37-39^{\circ F}$ & $25-33 \%$ & $25.6$ in & $7-18$ mph (E) & $0.0$ in & 15:20-37:10 \\
		\hline $24$th June & $72-77^{\circ F}$ & $43-50^{\circ F}$ & $31-39 \%$ & $25.6$ in & $5-14$ mph (ENE) & $0.0$ in & 15:20-37:10 \\
		\hline $26$th June & $68-81^{\circ F}$ & $46-50^{\circ F}$ & $30-44 \%$ & $25.6$ in & $2-18$ mph (ESE) & $0.0$ in & 15:20-37:10 \\
		\hline $28$th June & $70-81^{\circ F}$ & $48-50^{\circ F}$ & $30-46 \%$ & $25.6$ in & $7-10$ mph (SE) & $0.0$ in & 15:40-37:00 \\
		\hline $30$th June & $88-91^{\circ F}$ & $48-55^{\circ F}$ & $24-35 \%$ & $25.5$ in & $7-14$ mph (ENE) & $0.0$ in & 15:40-37:00 \\
		\hline $2$th July & $88-91^{\circ F}$ & $43-52^{\circ F}$ & $21-29 \%$ & $25.7$ in & $18-20$ mph (ENE) & $0.0$ in & 15:50-37:00 \\
		\hline $4$th July & $88-91^{\circ F}$ & $39-43^{\circ F}$ & $16-20 \%$ & $25.6$ in & $18-20$ mph (E) & $0.0$ in & 16:00-37:20 \\
		\hline $7$th July & $70-75^{\circ F}$ & $61-70^{\circ F}$ & $65-73 \%$ & $25.5$ in & $12-18$ mph (ENE) & $0.0$ in & 16:20-37:20 \\
		\hline
	\end{tabular}
	\caption{The weather conditions at which the photos were taken. $\alpha$ is the angle of the sunlight with respect to the normal vector of Earth.}
	\label{tab:conditions1}
\end{table*}

\begin{figure*}[t]
	\centering
	\begin{subfigure}{0.30\textwidth}\includegraphics[width=\textwidth]{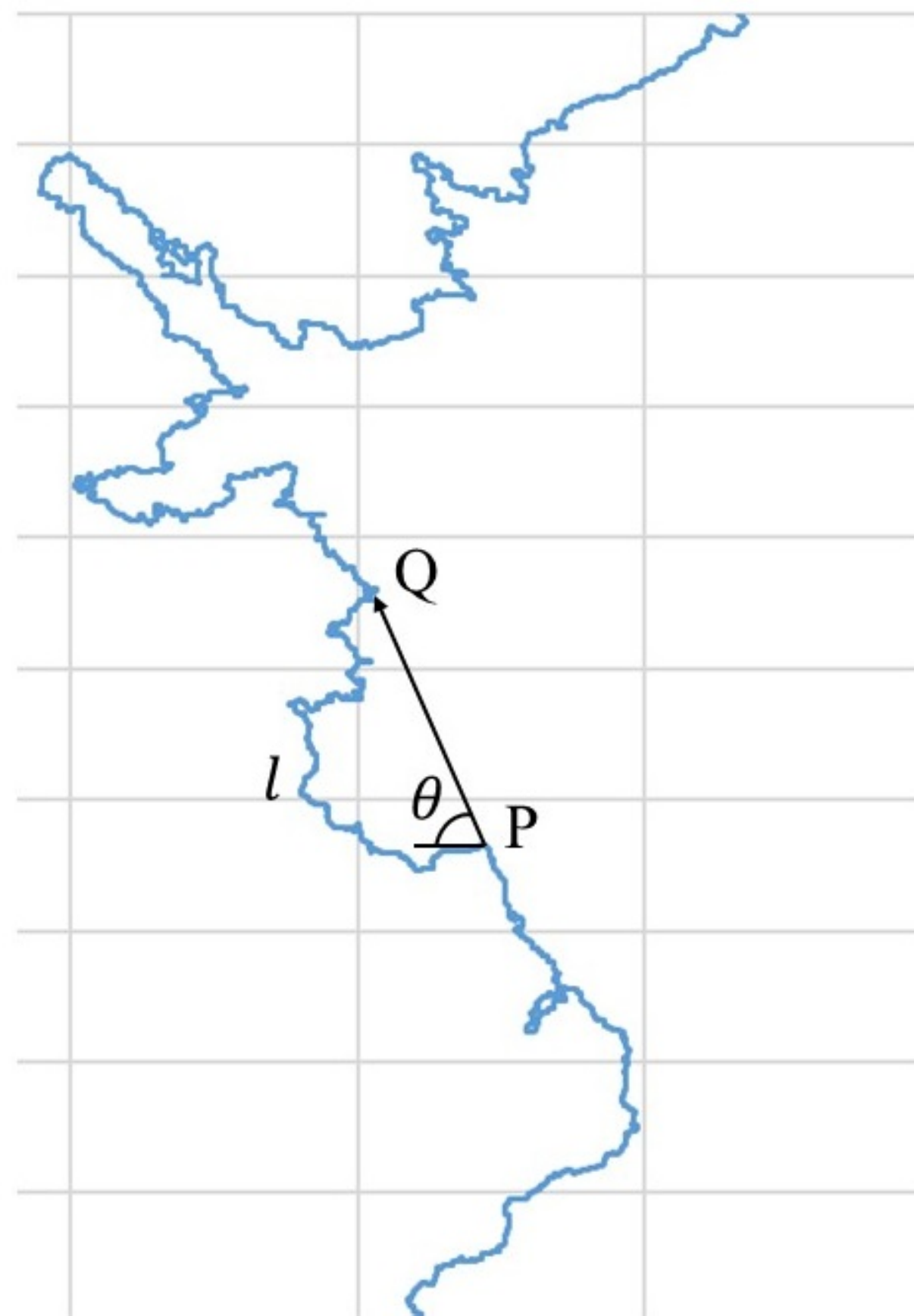}
		\caption{}
		\label{fig:sWA}
	\end{subfigure}
	\begin{subfigure}{0.30\textwidth}\includegraphics[width=\textwidth]{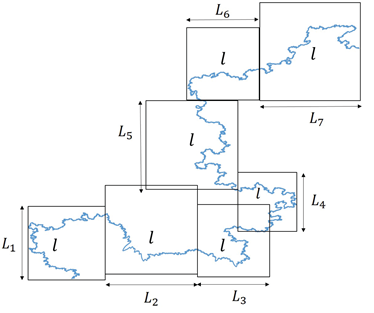}
		\caption{}
		\label{fig:BoxC}
	\end{subfigure}
	\begin{subfigure}{0.30\textwidth}\includegraphics[width=\textwidth]{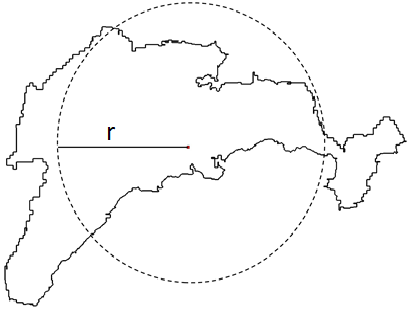}
		\caption{}
		\label{fig:gyration}
	\end{subfigure}
	\caption{(Color online) (a) Illustration of the winding angle statistics of random curves. (b) The box-counting method defining the quantities $l$ and $L$. (c) The $l-r$ scaling method for extracting the closed loops.}
	\label{fig:method_FD}
\end{figure*}

For open traces, the winding angle statistics of open traces is schematically shown in Fig.~\ref{fig:sWA}, in which one point is fixed on the curve, and another moves along the curve, and $\theta$ is defined by the angle between the straight line between them and the local slope. The fractal dimensions have been calculated by two different methods shown in Figs.~\ref{fig:BoxC} and~\ref{fig:gyration} sandbox (for BoCLs) and $l-r$ scaling (for BuCLs) respectively. For open traces we used the relation $l \sim \langle L \rangle^{D_f}$ in which $l$ is the length of the curve in a box of length $L$, which is related to the end-to-end exponent ($\nu$) via the relation $\nu =\frac{1}{D_f}$. To define the latter, if $R_i$ is the end-to-end Euclidean distance between the starting point and the $i$th point of the curve, then $\nu$ is defined by $\langle R^2 \rangle \sim N^{2\nu}$. For closed traces the fractal dimension is calculated using the relation $\left\langle \log l\right\rangle =D_f\left\langle \log r\right\rangle $ where $r$ represents the gyration radius as defined below:
\begin{equation}
r^2 = \frac{1}{l}\sum_{i=1}^{l} (\vec{r}_i - \vec{r}_{\text{com}})^2,
\end{equation}
where $ \vec{r}_{\text{com} } = \frac{1}{l}\sum_{i=1}^{l} \vec{r}_i$ is the center of mass of the closed trace.\\

\subsection{SLE}
Schramm-Loewner evolution (SLE) theory describes the critical behavior of 2D statistical models by focusing on their geometrical features such as their interfaces. SLE is the candidate to analyze these random curves by classifying them to one-parameter classes SLE$_{\kappa}$. There are three kinds of SLE; chordal SLE in which the random curve starts from zero and ends at infinity, dipolar SLE in which the curve starts from the boundary, and ends also at the boundary and radial SLE in which the curve starts from the boundary and ends in the bulk. In this paper we deal with chordal and dipolar SLEs.\\

Let us denote the upper half-plane by $H$ and $\gamma_t$ as the SLE trace grown up to time $t$. SLE$_{\kappa}$ is a growth process defined via conformal maps which are solutions of the stochastic Loewner's equation:
\begin{equation}
\partial_{t}g_{t}(z)=\frac{2}{g_{t}(z)-\xi_{t}},
\label{Loewner}
\end{equation}
in which the initial condition is $g_{t}(z)=z$  and the driving function $\xi_{t}$ is proportional to a one dimensional Brownian motion $B_{t}$ i.e. $\xi_{t}=\sqrt{\kappa}B_{t}$ in which $\kappa$ is the diffusivity parameter. $\tau_{z}$ is defined as the time for which for fixed $z$, $g_{t}(z)=\xi_{t}$. Although a trace cannot intersect itself, in the continuum limit where the lattice constant tends to zero, it might touch itself (in the dense phase $4\leq \kappa\leq 8$), so that some points that are not on the curve are separated from infinity, meaning that they are not reachable from infinity without crossing the SLE trace. The union of the set of such points, together with the SLE trace itself is called \textit{hull} $K_t$. Another way to define this is the complement of the connected component of $\overline{\lbrace z\in H:\tau_{z}\leq t \rbrace}$, which includes infinity, itself denoted by $H_{t}:=H\backslash{K_{t}}$. The map $g_{t}(z)$ is well-defined up to time $\tau_z$. This map is the unique conformal mapping $H_{t}\rightarrow{H}$ with $g_{t}(z)=z+\frac{2t}{z}+O(\frac{1}{z^{2}})$ as $z\rightarrow{\infty}$ known as hydrodynamical normalization. \\
\subsection{The discretization algorithm}
Consider a curve that starts from the origin and ends at a point on real-axis ($x_{\infty}$). The function $h_t\equiv \phi o G_t o \phi^{-1}$ describes chordal SLE in which $\phi(z)\equiv \frac{x_{\infty}z}{x_{\infty-z}}$ sends the end point of the curve to infinity, and $o$ represents the composition of maps. The equation governing $G_t(z)$ is
\begin{equation}
\partial_tG_t(z)=\frac{2}{\phi'(G_t)\left[ \phi(G_t)-\zeta_t\right] }
\end{equation}
whose solution for piecewise constant $\zeta$ is:
\begin{equation}
\begin{split}
&G_t(z)=x_{\infty}\frac{P(\zeta,x_{\infty},z)}{Q(\zeta,x_{\infty},z)}\\
& P(\zeta,x_{\infty},z)\equiv \eta x_{\infty} (x_{\infty}-z)+Q(\zeta,x_{\infty},z)\\
& Q(\zeta,x_{\infty},z)\equiv \sqrt{x_{\infty}^4(z-\eta)^2+4t(x_{\infty}-z)^2(x_{\infty}-\eta)^2}\\
&\eta \equiv \phi^{-1}(\zeta)
\end{split}
\end{equation}
To extract the driving function it is necessary to uniformize the discrete curve {$z_0 = 0,z_1 ,z_2 ,...,z_N $} (with length $l = N$) step by step.  To uniformize the first point, we consider $\zeta$ to be (piecewise) constant in the interval $(0,t_1)$ by the factors $\eta_0 = \phi^{-1}(\zeta_0)$ and $t_1=\frac{(\text{Im}(z_1))^2x_{\infty}^4}{4\left[ (\text{Re} z_1-x_{\infty})^2+ \text{Im}(z_1))^2\right]^2}$. Next the transformed point is chosen to be uniformized.\\
Apparently not all traces start and end on the same boundary of the system. Using the rotation shown in Fig.2 of the paper, we identify the straight line connecting the start and end point of the trace, and then cut the trace at the crossing points. \\

As a final point, we write here the explicit expression for the discretized Laplacian

\begin{equation}
\begin{split}
\nabla^2 V_i(x,y,z) &=\frac{1}{6} [ V_i(x+1,y,z)+V_i(x-1,y,z)+\\
&V_i(x,y+1,z) +V_i(x,y-1,z)+\\
&V_i(x,y,z+1) +V_i(x,y,z-1)-\\
&6 V_i(x,y,z)],
\end{split}
\end{equation}

and the third term (for $i=x$) is: 
\begin{equation}
\begin{split}
\nabla(\nabla.\vec{V})_x &= \frac{1}{2} [V_x(x+1,y,z)+V_x(x-1,y,z)-\\ &2V_x(x,y,z)]+\frac{1}{4}[V_y(x+1,y+1,z)+\\
&V_y(x-1,y-1,z) -V_y(x-1,y+1,z)-\\
&V_y(x+1,y-1,z)+V_z(x+1,y,z+1)+\\
&V_z(z-1,y,z-1)-V_z(x-1,y,z+1) -\\
& V_z(x+1,y,z-1)].
\end{split}
\end{equation}
The $y$ and $z$ components are calculated in a similar way. Concerning the definition of weights, let us consider the lattice point $(x,y,z)$ whose content are transported to a new position $(x +V_x,y +V_y,z+V_z ) =(i +\delta x , j + \delta y , k +\delta z ) $ after updating the velocity, where $i$, $j$ and $k$ are the integer part and  $\delta_x$,  $\delta_y$ and $\delta_z$ are the fractional portions. The adjacent sites are $(i\pm1,j,k)$, $(i,j\pm1,k)$, $(i,j,k\pm1)$, $(i\pm1,j\pm1,k)$, $(i\pm1,j,k\pm1)$, $(i, j\pm1, k\pm1)$ and $(i\pm 1,j\pm 1,k \pm 1)$, and  $\delta x$ as the distance between the new position and site $(i, j, k)$ in the $x$ direction, for the $y$ and $z$ directions we have $\delta y$ and $\delta z$. The weights are distributed between these $27$ sites (each site has $26$ adjacent sites), i.e. the variables are updated according to the weights at the sites. For the positive velocities ($\delta x,\delta y,\delta z>0$, for which only eight sites should be considered) the weights are $(1-\delta x)(1-\delta y)(1-\delta z)$, $\delta x(1-\delta y)(1-\delta z)$, $ (1-\delta x)\delta y (1-\delta z)$, $(1-\delta x)(1-\delta y)\delta z$,  $ \delta x\delta y(1-\delta z)$,  $\delta x(1-\delta y)\delta z$, $(1-\delta x)\delta y\delta z$, and $\delta x\delta y\delta z$, respectively.


\bibliography{refs}

\begin{thebibliography}{0}%
\makeatletter
\providecommand \@ifxundefined [1]{%
 \@ifx{#1\undefined}
}%
\providecommand \@ifnum [1]{%
 \ifnum #1\expandafter \@firstoftwo
 \else \expandafter \@secondoftwo
 \fi
}%
\providecommand \@ifx [1]{%
 \ifx #1\expandafter \@firstoftwo
 \else \expandafter \@secondoftwo
 \fi
}%
\providecommand \natexlab [1]{#1}%
\providecommand \enquote  [1]{``#1''}%
\providecommand \bibnamefont  [1]{#1}%
\providecommand \bibfnamefont [1]{#1}%
\providecommand \citenamefont [1]{#1}%
\providecommand \href@noop [0]{\@secondoftwo}%
\providecommand \href [0]{\begingroup \@sanitize@url \@href}%
\providecommand \@href[1]{\@@startlink{#1}\@@href}%
\providecommand \@@href[1]{\endgroup#1\@@endlink}%
\providecommand \@sanitize@url [0]{\catcode `\\12\catcode `\$12\catcode
  `\&12\catcode `\#12\catcode `\^12\catcode `\_12\catcode `\%12\relax}%
\providecommand \@@startlink[1]{}%
\providecommand \@@endlink[0]{}%
\providecommand \url  [0]{\begingroup\@sanitize@url \@url }%
\providecommand \@url [1]{\endgroup\@href {#1}{\urlprefix }}%
\providecommand \urlprefix  [0]{URL }%
\providecommand \Eprint [0]{\href }%
\providecommand \doibase [0]{http://dx.doi.org/}%
\providecommand \selectlanguage [0]{\@gobble}%
\providecommand \bibinfo  [0]{\@secondoftwo}%
\providecommand \bibfield  [0]{\@secondoftwo}%
\providecommand \translation [1]{[#1]}%
\providecommand \BibitemOpen [0]{}%
\providecommand \bibitemStop [0]{}%
\providecommand \bibitemNoStop [0]{.\EOS\space}%
\providecommand \EOS [0]{\spacefactor3000\relax}%
\providecommand \BibitemShut  [1]{\csname bibitem#1\endcsname}%
\let\auto@bib@innerbib\@empty
\end{thebibliography}%


\begin{thebibliography}{1}
\bibitem{Mandelbrot} Mandelbrot, Benoit B. The fractal geometry of nature. Vol. 173. New York: WH freeman, (1983).
\bibitem{Lovejoy} Lovejoy, S. Area-perimeter relation for rain and cloud areas. \textit{Science} \textbf{216}(4542), 185-187 (1982).
\bibitem{Austin} Austin, P.H., Baker, M.B., Blyth, A.M. \& Jensen, J.B. Small-scale variability in warm continental cumulus clouds. \textit{Journal of the atmospheric sciences} \textbf{42}(11), 1123-1138 (1985).
\bibitem{Chatterjee} Chatterjee, R.N., Ali, K. and Prakash, P. Fractal dimensions of convective clouds around Delhi. \textit{Indian Journal of Radio \& Space physics} \textbf{23} 189-192 (1994).
\bibitem{Savigny} von Savigny, C., Brinkhoff, L.A., Bailey, S.M., Randall, C.E. \& Russell, J.M. First determination of the fractal perimeter dimension of noctilucent clouds. \textit{Geophysical Research Letters} \textbf{38}(2) (2011).
\bibitem{Malinowski} Malinowski, S.P. \& Zawadzki, I. On the surface of clouds. \textit{Journal of the atmospheric sciences} \textbf{50}(1), 5-13 (1993).
\bibitem{Batista} Batista‐Tomás, A.R., Diaz, O., Batista‐Leyva, A.J. \& Altshuler, E. Classification and dynamics of tropical clouds by their fractal dimension. \textit{Quarterly Journal of the Royal Meteorological Society} \textbf{142}(695), 983-988 (2016).
\bibitem{Madhushani} Madhushani, K.N.R.A.K. \& Sonnadara, D.U.J. Fractal analysis of cloud shapes, In \textit{Proceedings of the Technical Sessions}, 28 59-64  (2012).
\bibitem{Joseph} Joseph, J.H. \& Cahalan, R.F. Nearest neighbor spacing of fair weather cumulus clouds. \textit{Journal of Applied Meteorology} \textbf{29}(8), pp.793-805 (1990).
\bibitem{Olsson} Olsson, J., Niemczynowicz, J. \& Berndtsson, R. Fractal analysis of high‐resolution rainfall time series.
\textit{Journal of Geophysical Research: Atmospheres} \textbf{98}(D12), 23265-23274 (1993).
\bibitem{Malinowski2} Malinowski, S.P., Leclerc, M.Y. \& Baumgardner, D.G. Fractal analyses of high-resolution cloud droplet measurements. \textit{Journal of the atmospheric sciences} \textbf{51}(3), 397-413 (1994).
\bibitem{Benner} Benner, T.C. \& Curry, J.A. Characteristics of small tropical cumulus clouds and their impact on the environment. \textit{Journal of Geophysical Research: Atmospheres} \textbf{103}(D22), 28753-28767 (1998).
\bibitem{Rodts} Rodts, S.M., Duynkerke, P.G. \& Jonker, H.J. Size distributions and dynamical properties of shallow cumulus clouds from aircraft observations and satellite data. \textit{Journal of the atmospheric sciences} \textbf{60}(16), 1895-1912 (2003).
\bibitem{Yano} Yano, J.I. \& Takeuchi, Y. The self-similarity of horizontal cloud pattern in the intertropical convergence zone. \textit{Journal of the Meteorological Society of Japan. Ser. II} \textbf{65}(4), 661-667 (1987).
\bibitem{Gotoh} Gotoh, K. \% Fujii, Y. A fractal dimensional analysis on the cloud shape parameters of cumulus over land. \textit{Journal of applied meteorology} \textbf{37}(10), 1283-1292 (1998).
\bibitem{Lovejoy2} Lovejoy, S. \& Schertzer, D. Multifractal analysis techniques and the rain and cloud fields from $10^{-3}$ to $10^6m$. In \textit{Non-Linear Variability in Geophysics} 111-144 Springer, Dordrecht (1991).
\bibitem{Lovejoy3} Lovejoy, S., Schertzer, D. \& Tsonis, A.A. Functional box-counting and multiple elliptical dimensions in rain. \textit{Science} \textbf{235}(4792), 1036-1038 (1987).
\bibitem{Cahalan} Cahalan, R.F. \& Joseph, J.H. Fractal statistics of cloud fields. \textit{Monthly weather review} \textbf{117}(2), 261-272 (1989).
\bibitem{Gabriel} Gabriel, P., Lovejoy, S., Schertzer, D. \& Austin, G.L. Multifractal analysis of resolution dependence in satellite imagery. \textit{Geophysical research letters} \textbf{15}(12), 1373-1376 (1988).
\bibitem{Lovejoy4} Lovejoy, S. \& Schertzer, D. Multifractals, universality classes and satellite and radar measurements of cloud and rain fields. \textit{Journal of Geophysical Research: Atmospheres} \textbf{95}(D3), 2021-2034 (1990).
\bibitem{Tessier} Tessier, Y., Lovejoy, S. \& Schertzer, D. Universal multifractals: Theory and observations for rain and clouds. \textit{Journal of Applied Meteorology} \textbf{32}(2), 223-250 (1993).
\bibitem{Pelletier} Pelletier, J.D. Kardar-Parisi-Zhang scaling of the height of the convective boundary layer and fractal structure of cumulus cloud fields. \textit{Physical review letters} \textbf{78}(13), 2672 (1997).
\bibitem{Sengupta} Sengupta, S.K., Welch, R.M., Navar, M.S., Berendes, T.A. \& Chen, D.W. Cumulus cloud field morphology and spatial patterns derived from high spatial resolution Landsat imagery. \textit{Journal of Applied Meteorology} \textbf{29}(12), 1245-1267 (1990).
\bibitem{Killen} Killen, R.M. \& Ellingson, R.G. The effect of shape and spatial distribution of cumulus clouds on longwave irradiance. Journal of the atmospheric sciences, \textbf{51}(14), 2123-2136 (1994).
\bibitem{Aida} Aida, M. Reflection of solar radiation from an array of cumuli. \textit{Journal of the Meteorological Society of Japan. Ser. II} \textbf{55}(2), 174-181 (1977).
\bibitem{Claussen} Claussen, M. On the radiative interaction in three-dimensional cloud fields. \textit{Contributions to Atmospheric Physics} \textbf{55} 158-169 (1982).
\bibitem{Parol} Parol, F., Buriez, J.C., Cretel, D. \& Fouquart, Y. February. The impact of cloud inhomogeneities on the Earth radiation budget: The 14 October 1989 ICE convective cloud case study. In \textit{Annales Geophysicae (Springer-Verlag)} \textbf{12}(2-3) 240-253 (1994).
\bibitem{Davies2} Davies, R.The effect of finite geometry on the three-dimensional transfer of solar irradiance in clouds. \textit{Journal of the Atmospheric Sciences} \textbf{35}(9), 1712-1725 (1978).
\bibitem{Welch} Welch, R.M. \& Wielicki, B.A. Reflected fluxes for broken clouds over a Lambertian surface. \textit{Journal of the Atmospheric Sciences}, \textbf{46}(10), 1384-1395 (1989).
\bibitem{Kite} Kite, A. The albedo of broken cloud fields. \textit{Quarterly Journal of the Royal Meteorological Society} \textbf{113}(476) 517-531 (1987).
\bibitem{Marshak} Marshak, A., Davis, A., Wiscombe, W. \& Cahalan, R. Radiative smoothing in fractal clouds. \textit{Journal of Geophysical Research: Atmospheres} \textbf{100}(D12), 26247-26261 (1995).
\bibitem{Cahalan3} Cahalan, R.F., Ridgway, W., Wiscombe, W.J., Bell, T.L. \& Snider, J.B. The albedo of fractal stratocumulus clouds. \textit{Journal of the Atmospheric Sciences} \textbf{51}(16), 2434-2455 (1994).
\bibitem{Miyazaki} Miyazaki, R., Yoshida, S., Dobashi, Y. \& Nishita, T. A method for modeling clouds based on atmospheric fluid dynamics. \textit{Proceedings Ninth Pacific Conference on Computer Graphics and Applications. Pacific Graphics} 2001 (pp. 363-372). IEEE (2001).
\bibitem{Kaneko} Kaneko, K. Simulating physics with coupled map lattices. In \textit{Formation, Dynamics And Statistics Of Patterns} \textbf{1} 1-54 (1990).
\bibitem{Wang} Wang, Y. \& Zocchi, G. Shape of fair weather clouds. \textit{Physical review letters} \textbf{104}(11), 118502 (2010).
\bibitem{Yano2} Yano, J.I., McWilliams, J.C. \& Moncrieff, M.W. Fractality in idealized simulations of large-scale tropical cloud systems. \textit{Monthly weather review} \textbf{124}(5), 838-848 (1996).
\bibitem{Richardson} Richardson, L.F. Atmospheric diffusion shown on a distance-neighbour graph.\textit{ Proceedings of the Royal Society of London. Series A, Containing Papers of a Mathematical and Physical Character} \textbf{110}(756), 709-737 (1926).
\bibitem{Hentschel} Hentschel, H.G.E. \& Procaccia, I. Relative diffusion in turbulent media: the fractal dimension of clouds. \textit{Physical Review A} \textbf{29}, 1461 (1984).
\bibitem{Siebesma} Siebesma, A.P. \& Jonker, H.J.J. Anomalous scaling of cumulus cloud boundaries. \textit{Physical review letters} \textbf{85}(1), 214 (2000).
\bibitem{Zhao} Zhao, M. \& Austin, P.H. Life cycle of numerically simulated shallow cumulus clouds. Part I: Transport. \textit{Journal of the atmospheric sciences} \textbf{62}(5), 1269-1290 (2005).
\bibitem{Schertzer} Schertzer, D. \& Lovejoy, S. Physical modeling and analysis of rain and clouds by anisotropic scaling multiplicative processes. \textit{Journal of Geophysical Research: Atmospheres} \textbf{92}(D8), 9693-9714 (1987).
\bibitem{Nagel} Nagel, K. \& Raschke, E. Self-organizing criticality in cloud formation? \textit{Physica A: Statistical Mechanics and its Applications}, \textbf{182}(4), 519-531 (1992).
\bibitem{Gardner} Gardner, G.Y. Visual simulation of clouds. \textit{Proceedings of the 12th annual conference on Computer graphics and interactive techniques} 297-304 (1985).
\bibitem{Cianciolo} Cianciolo, M.E., Cumulus cloud scene simulation modeling using fractals and physics. In \textit{Digital Image Processing and Visual Communications Technologies in the Earth and Atmospheric Sciences II} (Vol. 1819, pp. 70-77). International Society for Optics and Photonics (1993).
\bibitem{Bouthors} Bouthors, A. and Neyret, F. Modeling clouds shape, \textit{EUROGRAPHICS} August (2004).
\bibitem{Lohmann} Lohmann, G.M., Hammer, A., Monahan, A.H., Schmidt, T. \& Heinemann, D. Simulating clear-sky index increment correlations under mixed sky conditions using a fractal cloud model. \textit{Solar Energy} \textbf{150}, 255-264 (2017).
\bibitem{Cai} Cai, C. \& Aliprantis, D.C. Cumulus cloud shadow model for analysis of power systems with photovoltaics. \textit{IEEE Transactions on Power Systems} \textbf{28}(4), 4496-4506 (2013).
\bibitem{Rys} Rys, F.S. and Waldvogel, A. Fractal shape of hail clouds. \textit{Physical review letters} \textbf{56}(7), 784 (1986).
\bibitem{Peters} Peters, O. and Neelin, J.D. Critical phenomena in atmospheric precipitation. \textit{Nature physics} \textbf{2}(6), 393-396 (2006).
\bibitem{Beyer} Beyer, H.G., Hammer, A., Luther, J., Poplawska, J., Stolzenburg, K. \& Wieting, P. Analysis and synthesis of cloud pattern for radiation field studies. \textit{Solar energy} \textbf{52}(5),379-390 (1994).
\bibitem{Lovejoy9} Lovejoy, S., Schertzer, D., Silas, P., Tessier, Y. \& Lavallée, D., The unified scaling model of atmospheric dynamics and systematic analysis of scale invariance in cloud radiances. In \textit{Annales Geophysicae} \textbf{11}, No. 2, pp. 119-127) (1993).
\bibitem{Lawler} Lawler, G.F., Schramm, O. \& Werner, W. Conformal invariance of planar loop-erased random walks and uniform spanning trees. In \textit{Selected Works of Oded Schramm} 931-987. Springer, New York, NY (2011).
\bibitem{Duplantier} Duplantier, B. \& Saleur, H. Winding-angle distributions of two-dimensional self-avoiding walks from conformal invariance. \textit{Physical review letters} \textbf{60}(23), 2343 (1988).
\bibitem{Wieland} Wieland, B. \& Wilson, D.B. Winding angle variance of Fortuin-Kasteleyn contours. \textit{Physical Review E} \textbf{68}(5), 056101 (2003).
\bibitem{Kondev} Kondev, J., Henley, C.L. \& Salinas, D.G. Nonlinear measures for characterizing rough surface morphologies. \textit{Physical Review E} \textbf{61}(1), 104 (2000).
\bibitem{Najafi0} Najafi, M.N., Moghimi-Araghi, \& S. and Rouhani, S. Observation of SLE ($\kappa,\rho$) on the critical statistical models. \textit{Journal of Physics A: Mathematical and Theoretical}, \textbf{45}(9), 095001 (2012).
\bibitem{Bauer1} Bauer, M., Bernard, D. \& Kytölä, K. LERW as an example of off-critical SLEs. \textit{Journal of Statistical Physics}, \textbf{132}(4), 721-754 (2008).
\bibitem{Dhar} Dhar, D. The abelian sandpile and related models. \textit{Physica A: Statistical Mechanics and its Applications}, \textbf{263}(1-4), 4-25 (1999).
\bibitem{Francesco} Francesco, P., Mathieu, P. \& Senechal, D. Conformal field theory. \textit{Springer Science \& Business Media} (2012).
\bibitem{Rohde1} Rohde, S. \& Schramm, O. Basic properties of SLE. In Selected Works of Oded Schramm 989-1030. Springer, New York, NY (2011).
\bibitem{Cardy} Cardy, J. SLE for theoretical physicists. \textit{Annals of Physics} \textbf{318}(1), 81-118 (2005).
\bibitem{Bernard2} Bernard, D., Boffetta, G., Celani, A. \& Falkovich, G. Conformal invariance in two-dimensional turbulence. \textit{Nature Physics} \textbf{2}(2),124-128 (2006).
\bibitem{Najafi2} Najafi, M.N., Moghimi-Araghi, S. \& Rouhani, S. Avalanche frontiers in the dissipative Abelian sandpile model and off-critical Schramm-Loewner evolution. \textit{Physical Review E} \textbf{85}(5), 051104 (2012).
\bibitem{Majumdar} Majumdar, S.N. \& Dhar, D. Equivalence between the Abelian sandpile model and the $q\rightarrow 0$ limit of the Potts model. \textit{Physica A: Statistical Mechanics and its Applications} \textbf{185}(1-4), 129-145 (1992).
\bibitem{Mahieu} Mahieu, S. \& Ruelle, P. Scaling fields in the two-dimensional abelian sandpile model. \textit{Physical Review E} \textbf{64}(6), 066130 (2001).
\bibitem{Najafi1} Najafi, M.N., Cheraghalizadeh, J., Lukovic, M. \& Herrmann, H.J. Geometry-induced nonequilibrium phase transition in sandpiles. \textit{Physical Review E} \textbf{101}(3), 032116 (2020).
\bibitem{Daryaei} Daryaei, E., Araujo, N.A.M., Schrenk, K.J., Rouhani, S. \& Herrmann, H.J. Watersheds are Schramm-Loewner evolution curves. \textit{Physical review letters} \textbf{109}(21), 218701 (2012).
\bibitem{Yanagita} Yanagita, T. \& Kaneko, K. Modeling and characterization of cloud dynamics. \textit{Physical Review Letters}, \textbf{78}(22), p.4297 (1997).
\bibitem{Yanagita2} Yanagita, T. \& Kaneko, K. Rayleigh-Bénard convection patterns, chaos, spatiotemporal chaos and turbulence. \textit{Physica D: Nonlinear Phenomena}, \textbf{82}(3), pp.288-313 (1995).
\bibitem{Blanz} Blanz, V. \& Vetter, T. \textit{Proceedings of the 26th annual conference on computer graphics and interactive techniques} (1999).
\bibitem{Cussler} Cussler, E.L. \& Cussler, E.L. Diffusion: mass transfer in fluid systems. \textit{Cambridge university press} (2009).
\end{thebibliography}
\end{document}